\documentclass[acmsmall]{acmart}
\AtBeginDocument{%
  \providecommand\BibTeX{{%
    Bib\TeX}}}

\usepackage{amsmath,amssymb,amsfonts}
\usepackage{algorithmic}
\usepackage{graphicx}
\usepackage{subcaption}
\usepackage{textcomp}
\usepackage{caption}
\usepackage{xcolor}
\usepackage{colortbl}
\usepackage{enumitem}
\usepackage{multirow}
\usepackage{url}
\usepackage{hyperref}
\usepackage{balance}
\usepackage[breakable]{tcolorbox}
\def\BibTeX{{\rm B\kern-.05em{\sc i\kern-.025em b}\kern-.08em
    T\kern-.1667em\lower.7ex\hbox{E}\kern-.125emX}}
\usepackage{xspace}
\usepackage[linesnumbered,ruled,vlined]{algorithm2e}
\usepackage{float}
\usepackage{booktabs,siunitx,threeparttable,multirow}
\definecolor{EffiSkelRow}{gray}{0.93} 
\newcommand{\bplus}[1]{{\bfseries #1}}                 
\newcommand{\eskc}[1]{\cellcolor{EffiSkelRow}{\bplus{#1}}} 
\newcommand{\eskm}[1]{\cellcolor{EffiSkelRow}{\bplus{#1}}} 
\newcommand{\best}[1]{\cellcolor{EffiSkelRow}{\bplus{#1}}}

\usepackage{longfbox}
\definecolor{ashgrey}{rgb}{0.7, 0.75, 0.71}
\definecolor{grey}{rgb}{0.6,0.6,0.6}

\definecolor{green1}{RGB}{200, 230, 200} 
\definecolor{green2}{RGB}{150, 210, 150} 
\definecolor{green3}{RGB}{100, 190, 100} 
\definecolor{green4}{RGB}{50, 160, 50}
\definecolor{green5}{RGB}{240, 250, 240} 

\newcommand{\mypara}[1]{\smallskip \noindent\textbf{#1.} \xspace}

\usepackage{siunitx}
\setcopyright{cc}
\setcctype{by}
\acmDOI{10.1145/3808194}
\acmYear{2026}
\acmJournal{PACMSE}
\acmVolume{3}
\acmNumber{FSE}
\acmArticle{FSE187}
\acmMonth{7}
\acmSubmissionID{fse26mainb-p2213-p}
\received{2026-02-08}
\received[accepted]{2026-03-24}

\begin{document}

\title{Chiseling Out Efficiency: Structured Skeleton Supervision for Efficient Code Generation}

\author{Yu Yu}
\orcid{0009-0002-7096-2750}
\affiliation{%
  \department{School of Computer Science and Artificial Intelligence}
  \institution{Shandong Normal University}
  \city{Jinan}
  \country{China}
}
\email{2023028044@stu.sdnu.edu.cn}

\author{Zhihong Sun}
\orcid{0009-0007-1387-3010}
\affiliation{%
  \institution{Nanjing University}
  \city{Nanjing}
  \country{China}
}
\email{zhihong.sun@smail.nju.edu.cn}

\author{Jia Li}
\orcid{0000-0002-5579-8852}
\affiliation{%
  \institution{Tsinghua University}
  \city{Beijing}
  \country{China}
}
\email{jia_li@mail.tsinghua.edu.cn}

\author{Yao Wan}
\orcid{0000-0001-6937-4180}
\affiliation{%
  \institution{Huazhong University of Science and Technology}
  \city{Wuhan}
  \country{China}
}
\email{wanyao@hust.edu.cn}

\author{Chuanyi Li}
\orcid{0000-0001-9270-5072}
\affiliation{%
  \institution{Nanjing University}
  \city{Nanjing}
  \country{China}
}
\email{lcy@nju.edu.cn}

\author{Hongyu Zhang}
\orcid{0000-0002-3063-9425}
\affiliation{%
  \institution{Chongqing University}
  \city{Chongqing}
  \country{China}
}
\email{hyzhang@cqu.edu.cn}

\author{Ruyun Wang}
\orcid{0000-0002-7113-3017}
\affiliation{%
  \institution{Institute of Information Engineering at Chinese Academy of Sciences}
  \city{Beijing}
  \country{China}
}
\email{wangruyun@iie.ac.cn}

\author{Tao Huang}
\orcid{0009-0009-6955-7417}
\affiliation{%
  \department{School of Computer Science and Artificial Intelligence}
  \institution{Shandong Normal University}
  \city{Jinan}
  \country{China}
}
\email{2022317095@stu.sdnu.edu.cn}

\author{Zhi Jin}
\orcid{0000-0003-1087-226X}
\affiliation{%
  \institution{Peking University}
  \city{Beijing}
  \country{China}
}
\email{zhijin@pku.edu.cn}

\author{Ge Li}
\orcid{0000-0002-5828-0186}
\affiliation{%
  \institution{Peking University}
  \city{Beijing}
  \country{China}
}
\email{lige@pku.edu.cn}

\author{Chen Lyu}
\orcid{0000-0002-5044-1459}
\authornote{Chen Lyu is the corresponding author.}
\affiliation{%
  \department{School of Computer Science and Artificial Intelligence}
  \institution{Shandong Normal University}
  \city{Jinan}
  \country{China}
}
\email{lvchen@sdnu.edu.cn}

\renewcommand{\shortauthors}{Yu et al.}

\begin{abstract}
\textit{Large Language Models} (LLMs) are capable of
generating syntactically correct and functionally complete programs, greatly streamlining software development.
However, recent studies reveal that these programs typically execute substantially slower than human-optimized counterparts.
Existing approaches to bridging this efficiency gap typically involve either iteratively optimizing code after generation or fine-tuning models on corpora of efficient code.
Yet, these methods expose the model to efficiency signals only by mimicking complete, optimized solutions, without explicitly encoding the structural code patterns essential for achieving high runtime performance.
Addressing this gap presents two core challenges: (1) extracting and representing latent, efficiency-oriented structural patterns embedded within complex syntax and control flows, and (2) effectively learning these patterns without destabilizing the semantic training of LLMs.
To tackle these challenges, we propose \textsc{EffiSkel}, an \textit{effi}ciency \textit{skel}eton-guided framework that explicitly extracts and learns efficiency skeletons—abstract, reusable structural patterns underpinning efficient code—by leveraging three complementary strategies: lexical analysis based on token-frequency saliency, syntactic analysis using similarity over Abstract Syntax Trees (ASTs), and dynamic line-level profiling of execution time.
These skeletons are integrated into a multi-task learning regime that jointly optimizes code generation and skeleton prediction, introducing an explicit inductive bias toward efficiency-aware code generation.
Experiments across multiple programming languages and benchmarks demonstrate that \textsc{EffiSkel} significantly enhances both functional correctness and efficiency, resulting on \textit{Mercury} with DeepSeek-Coder (6.7B) a +11.11\% (vs. \textsc{EffiCoder}) and +3.71\% (vs. \textsc{CodeDPO}) higher Efficiency Ratio (ER), and a +0.36 (vs. \textsc{EffiCoder}) and +0.22 (vs. \textsc{CodeDPO}) increase in Average Speedup (AS).
These results highlight the effectiveness of explicitly modeling efficiency skeletons in improving the runtime performance of code generated by LLMs.
\end{abstract}




\maketitle

\section{Introduction}

\textit{Large Language Models} (LLMs)—such as GPT-4o~\cite{openai2024gpt4o}, DeepSeek-R1~\cite{guo2025deepseek}, Code Llama~\cite{roziere2023code}, and StarCoder2~\cite{lozhkov2024starcoder}—have demonstrated remarkable abilities to generate syntactically correct and functionally complete programs, greatly streamlining software development workflows.  
However, their execution efficiency remains insufficient: recent evaluations reveal that GPT-4–generated programs are on average \textbf{1.69$\times$} slower than human-optimized code, and in extreme cases, performance degradation can reach up to \textbf{45.49$\times$}~\cite{niu2024evaluating, huang2024effilearner}.  
This widening gap between correctness and performance severely limits their application in large-scale industrial environments.

Efficient code is essential for green computing and sustainable software engineering, directly reducing energy consumption and carbon emissions~\cite{cappendijk2024generating, peng2024large, shi2024efficient, feng2024llmeffichecker, liu2024evaluating}.  
With the rapidly expanding global demand for digital services, optimizing runtime efficiency has transitioned from a desirable attribute to a fundamental necessity.



Existing approaches to bridge this efficiency gap primarily follow two directions: efficient code optimization and direct efficient code generation.
Efficient code optimization feeds the original code back into the prompt and iterates: execute the program to collect a runtime profile, locate hotspots, and re-prompt the model to rewrite the identified parts (Effi-Learner~\cite{huang2024effilearner}, LLM4EFFI~\cite{ye2025llm4effi}, Afterburner~\cite{du2025afterburner}); it can yield faster code but remains tied to the initial implementation, adds inference-time compute, and often converges to local optima.
Direct efficient code generation provides a complementary route. \textsc{EffiCoder}~\cite{huangefficoder} performs efficiency-aware supervised fine-tuning on curated corpora so the model emits efficient programs by construction; \textsc{CodeDPO}~\cite{zhang2024codedpo} applies Direct Preference Optimization over pairs of efficient and inefficient implementations to align generation toward efficiency while preserving correctness.
However, \textsc{EffiCoder} and \textsc{CodeDPO} learns about efficiency exclusively from the raw programs. It does not explicitly utilize structural code patterns—\emph{efficiency skeletons} such as optimal algorithmic selections, efficient control flows, and effective memory management—that truly govern runtime performance. Deprived of this guidance, the model merely mimics surface token sequences and fails to encode why certain transformations accelerate execution. We believe that supplying those skeletons as fine-grained supervision would provide a powerful teaching signal, helping the model reason about, and consistently reproduce the principles of efficient coding.

Motivated by this, a natural question arises:
\emph{Can we explicitly model these efficiency skeletons to guide LLMs beyond mere outcome imitation and toward generating inherently efficient code?}
Addressing this question requires shifting from supervision based on pre-existing optimized code examples to guidance derived from explicit efficiency skeleton supervision. From our investigation, this perspective shift introduces two fundamental research challenges: {\bfseries Challenge 1: Representing and Extracting Efficiency Skeletons.} Efficiency-critical structures are implicitly embedded within complex syntax and intricate control flows. Extracting abstract yet expressive efficiency skeletons, without compromising fine-grained performance-critical semantics, presents significant complexity. {\bfseries Challenge 2: Learning Efficiency Skeletons Effectively.} Direct end-to-end training of LLMs inherently entangles efficiency-critical structural signals with semantic information, making it challenging to encode efficiency skeletons without destabilizing model convergence.

\begin{figure}[t]
  \centering
  \includegraphics[width=.7\linewidth]{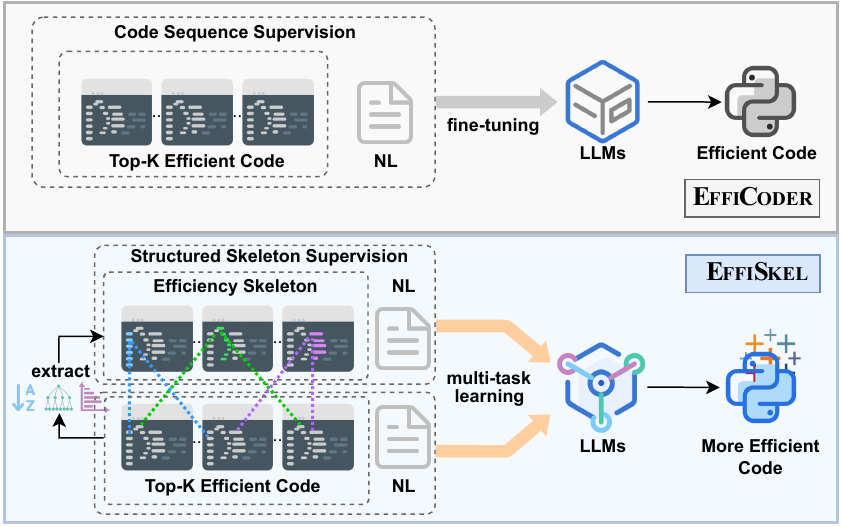}
  \caption{\textsc{EffiCoder} vs.\ \textsc{EffiSkel}.  
           \textsc{EffiCoder} relies on code sequence supervision;  
           \textsc{EffiSkel} employs structured skeleton supervision in a multi-task learning framework, first extracting efficiency skeleton and then jointly training for skeleton prediction and code generation.}
  \label{fig:high-level}
\end{figure}

To effectively address these challenges, we propose \textsc{EffiSkel}, an \underline{Effi}ciency \underline{Skel}eton-guided framework that explicitly captures and leverages efficiency-critical structures during code generation.  
As illustrated in Figure~\ref{fig:high-level}, unlike conventional methods (upper part) that only rely on fine-tuning on efficient code, \textsc{EffiSkel} explicitly extracts efficiency skeletons from efficient programs, using these skeletons as structural supervision signals during training. Specifically, to address \textit{Challenge 1}, we first construct the \textit{APPS+EFFI} dataset, an efficient subset of \textit{APPS}~\cite{hendrycks2021measuring} containing functionally correct and efficient code mined via performance profiling. Building on this dataset, we develop a multi-layered skeleton extraction pipeline integrating static and dynamic analysis. We introduce three complementary extraction strategies: (1) lexical analysis via token-frequency saliency (common tokens), (2) syntactic based on abstract syntax tree (AST) similarity (shared structures), and (3) dynamic execution profiling at line-level granularity (performance hotspots). Collectively, frequent lexical idioms pinpoint that recur across fast solutions~\cite{yang2024streamlining}; similar AST fragments capture the invariant control-flow skeleton of efficient algorithms~\cite{song2024revisiting}; and profiler-validated hot blocks isolate the irreducible work that dominates runtime~\cite{zhao2024easyview}, these strategies capture a comprehensive set of performance-relevant patterns. To resolve \textit{Challenge 2}, we incorporate skeleton prediction as an auxiliary learning objective within a structure-aware multi-task training paradigm, jointly optimizing code generation and skeleton inference. This explicit structural supervision ensures models effectively encode efficiency-related patterns alongside semantic content, thereby enhancing training stability and overall runtime performance awareness.

We empirically validate \textsc{EffiSkel} under a multi-task learning framework and systematically evaluate its capability for efficient code generation across five mainstream LLMs. Functional correctness is measured by Pass@K, while efficiency is assessed using ER and AS (See Section ~\ref{ER}). Experiments are conducted on seven datasets—\textit{Mercury}, \textit{ENAMEL}, \textit{APPS}, \textit{EffiBench},  \textit{HumanEval-X} (Java), \textit{DevEval} and \textit{CoderEval}—culminating in 45 distinct experimental configurations (5 LLMs × 3 skeleton extraction strategies × 3 training methods). Results across languages and benchmarks show that \textsc{EffiSkel} consistently improves both functional correctness and runtime efficiency under diverse settings. For example, on \textit{Mercury} with DeepSeek\mbox{-}Coder (6.7B), relative to prompting-based methods we obtain \textbf{+16.67\%} (vs.\ \textsc{Instruct}) and \textbf{+14.82\%} (vs.\ \textsc{LLM4EFFI}) gains in ER, together with \textbf{+0.47} (vs.\ \textsc{Instruct}) and \textbf{+0.43} (vs.\ \textsc{LLM4EFFI}) increases in AS; against stronger training-based baselines, \textsc{EffiSkel} further achieves \textbf{+11.11\%} (vs.\ \textsc{EffiCoder}) and \textbf{+3.71\%} (vs.\ \textsc{CodeDPO}) in ER, with \textbf{+0.36} (vs.\ \textsc{EffiCoder}) and \textbf{+0.22} (vs.\ \textsc{CodeDPO}) improvements in AS. Meanwhile, functional correctness (Pass@K) is further improved.

In summary, the main contributions of this paper are as follows:

\begin{itemize}
\item \textbf{Conceptual Innovation.} 
We propose the concept of an \emph{efficiency skeleton} to highlight structural aspects that strongly influence code efficiency. While efficiency also depends on external factors (e.g., hardware or compilers), we focus on structural properties as they offer actionable, learnable signals for LLMs. By using these patterns as explicit supervision—rather than relying solely on code examples—we guide models to encode algorithmic best practices and performance-aware programming more effectively.

\item \textbf{Technical Advances.} We propose three complementary strategies to systematically extract representative efficiency skeletons. Moreover, we develop a structure-aware multi-task learning framework that jointly optimizes skeleton prediction and code generation, effectively embedding efficiency insights directly into LLM training.

\item \textbf{Empirical Validation.} We introduce the \textit{APPS+EFFI} benchmark, explicitly focusing on efficiency-critical code generation tasks, and demonstrate through extensive experiments that \textsc{EffiSkel} achieves significant improvements in runtime efficiency across multiple programming languages and benchmarks.
    
\end{itemize}

\section{Motivation}

\subsection{A Motivating Example}

\begin{figure}[!h]
    \centering
    \includegraphics[width=.7\columnwidth]{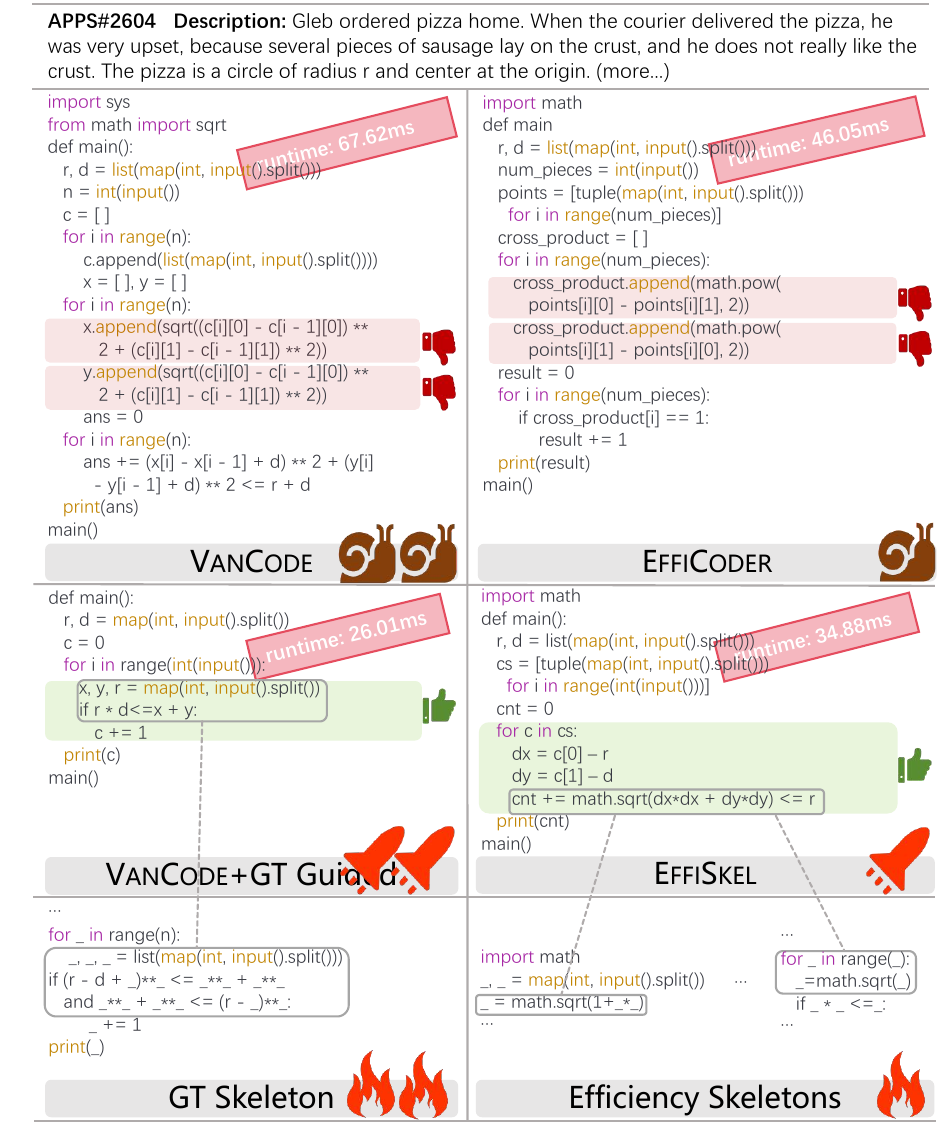}
    \caption{A representative task from the \textit{APPS} test set, comparing code generated by CodeT5 (770M): fine-tuned on vanilla code (\textsc{VanCode}),  fine-tuned on efficient code (\textsc{EffiCoder}), \textsc{VanCode} further guided by ground-truth skeletons (\textsc{VanCode+GT Guided}) and trained with our proposed skeleton-guided framework (\textsc{EffiSkel}). Red boxes indicate inefficient patterns; green boxes highlight efficient strategies.}
    \label{fig: motivating}
    \vspace{-4mm}
\end{figure}

Figure~\ref{fig: motivating} illustrates a representative problem from the \textit{APPS} test set, which involves determining if given points lie within certain geometric boundaries defined by circles. We compare code generated by four distinct methods: \textsc{VanCode} (top-left), \textsc{EffiCoder} (top-right), \textsc{VanCode+GT Guided} (middle-left), and our proposed \textsc{EffiSkel} (middle-right).

The \textsc{VanCode} method, fine-tuned solely on vanilla code, adopts code involving redundant floating-point operations, such as repeated calls to expensive \verb|sqrt| functions, leading to considerable runtime inefficiencies. Even the \textsc{EffiCoder} method, fine-tuned specifically on efficiency-oriented code, still exhibits superficial improvements, maintaining unnecessary duplicate computations and suboptimal data structure usage, thus providing only minor practical performance gains.

In contrast, the \textsc{VanCode+GT Guided} approach explicitly leverages the ground-truth (GT) efficiency skeletons (bottom-left) extracted directly from the test samples by the TAS method, please refer to Section~\ref{sec:extract}. The GT skeleton specifies integer-based distance comparisons, such as \verb|if(r - d + _)**_ <= _**_ + _**_|, and explicitly guides the model to avoid expensive floating-point operations, such as \verb|if r*d <= x + y:|. By using these specific contexts provided by the GT skeleton, the model can effectively leverage  such a skeleton through in-context learning, thereby significantly improving the efficiency of the generated code.

However, in practical scenarios, obtaining such GT skeletons is unrealistic. As a practical alternative, our \textsc{EffiSkel} framework learns general efficiency skeletons (bottom-right) from training set, rather than relying solely on fine-tuned efficient code examples. \textsc{EffiSkel} effectively captures critical optimization strategies, such as integer-based comparisons and reduced floating-point operations, significantly enhancing generated code efficiency. For example,  \verb|_ = math.sqrt(1+_*_)| and \verb|_math.squrt(_)| in the efficiency skeletons can prompt the \textsc{EffiSkel} model to generate efficient code like \verb|cnt += math.sqrt(dx*dx + dy*dy) <= r|. Although \textsc{EffiSkel} may not fully reach the ideal performance of \textsc{VanCode+GT Guided}, it substantially surpasses \textsc{EffiCoder} by eliminating redundant computations and unnecessary intermediate data structures, leading to notable runtime improvements. We fill the non-skeleton part with \verb|<MASK>|, which is represented by \verb|_| here.

This motivating example reveals a key limitation of existing fine-tuning methods: training on efficient code alone often leads the model to mimic surface token patterns without understanding why certain transformations improve execution efficiency. In contrast, structured skeleton guidance provides fine-grained supervision that helps the model encode performance-critical strategies and consistently reproduce them across diverse tasks.

\subsection{Key Idea}

While efficiency-oriented code skeletons hold great promise for guiding LLMs toward more performant code structures, the practical challenge remains that ground-truth skeletons are rarely available at inference time. Models trained without explicit skeleton supervision struggle to encode reusable efficiency patterns, ultimately limiting their runtime performance.

To address this, we propose a novel multi-task learning framework that explicitly integrates code generation with efficiency skeleton  modeling. Our key innovation lies in introducing skeleton prediction as an auxiliary learning task, combined with structure-aware generation and position-sensitive masking. By aligning semantic and structural objectives through shared representations, our framework enables models to encode critical performance heuristics, achieving generalization to efficient coding strategies—even in the absence of ground-truth skeletons during inference.

\begin{table}[!h]
\centering
\footnotesize

\caption{Comparison of CodeT5 (770M) under three training methods.
Cells report Efficiency Ratio (ER) when the row method is compared against the column method; green indicates gains and red indicates declines.}
\label{tab:table9}
\renewcommand{\arraystretch}{1.08}
\begin{threeparttable}
\begin{tabular}{lccc}
\toprule
\textbf{CodeT5} & \textbf{\textsc{VanCode}} & \textbf{\textsc{EffiCoder}} & \textbf{\textsc{EffiSkel}} \\
\midrule
\textbf{\textsc{VanCode}}   & \textemdash & \cellcolor{red!40} 38.11\% & \cellcolor{red!60} 32.95\% \\
\textbf{\textsc{EffiCoder}} & \cellcolor{green!40} 61.89\% & \textemdash & \cellcolor{red!20} 45.91\% \\
\textbf{\textsc{EffiSkel}}  & \cellcolor{green!60} 67.05\% & \cellcolor{green!20} 54.09\% & \textemdash \\
\addlinespace[2pt]
\midrule
\textbf{Pass@10} & 2.29 & 2.24 & \textbf{2.67} \\
\bottomrule
\end{tabular}
\end{threeparttable}
\vspace{-0.6em}
\end{table}

\subsection{Feasibility Analysis}

To empirically verify the feasibility of our approach, we conduct preliminary experiments with the CodeT5 (770M)~\cite{wang2021codet5} model. For training, we leverage our \textit{APPS+EFFI} dataset containing efficient solutions, constructing comparative baselines by sampling equal-sized vanilla training sets from \textit{APPS}. To streamline evaluations, we randomly select 1,000 problems from the \textit{APPS} test set. As summarized in Table~\ref{tab:table9}, the model trained with explicit skeleton supervision (\textsc{EffiSkel}) significantly outperforms the results of fine-tuning on vanilla code (\textsc{VanCode}) and fine-tuning on efficient code (\textsc{EffiCoder}), measured by Efficiency Ratio (ER) (See Section ~\ref{ER}).

These results confirm the feasibility and effectiveness of employing structured skeleton supervision for efficiency-oriented code generation, laying a robust foundation for training LLMs to encode and generalize critical performance patterns.
\section{Approach}

We now present the detailed formulation and training procedure of \textsc{EffiSkel}, a framework designed to enhance code generation efficiency through structure-aware supervision. An overview of the full pipeline is shown in Figure~\ref{fig: EffiSkel}, which consists of three key stages: data profiling, skeleton extraction, and model optimization.

\subsection{Dataset Profiling via Execution Time Measurement}
\label{sec:data}

\begin{figure*}
    \centering
    \includegraphics[width=\linewidth]{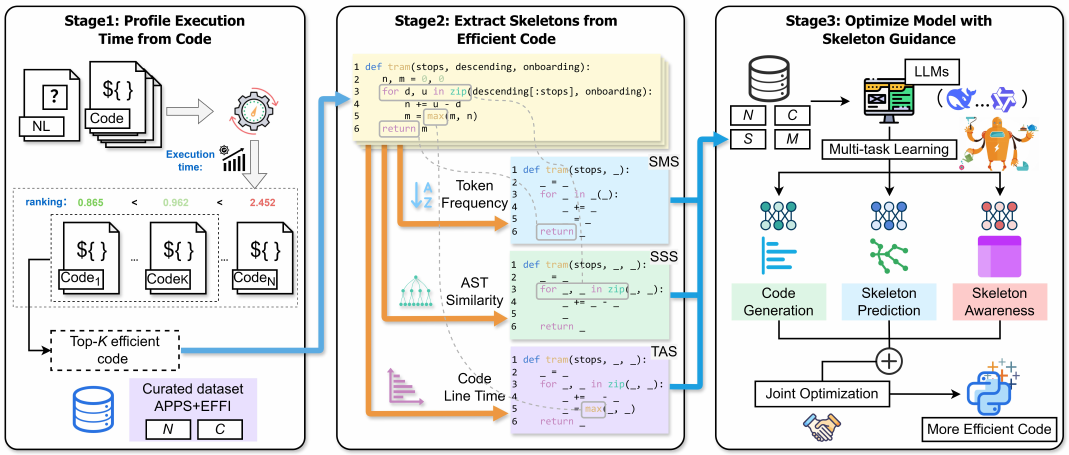}
    \caption{Overview of the EffiSkel framework, which consists of three stages: (1) profile and rank candidate programs by executing each on \textit{APPS} training set, selecting the top-$K$ efficient code to form the \textit{APPS+EFFI} subset, (2) extract efficiency skeletons from these efficient solutions via three methods—token frequency (SMS), AST similarity (SSS), and line-level execution profiling (TAS), and (3) optimize the model via multi-task learning on the four-tuple (N, C, S, M)—natural language description, efficient code, its skeleton, and the corresponding mask, jointly training on code sequence supervision, skeleton prediction, and skeleton awareness objectives, thereby improving both functional correctness and runtime efficiency.}
    \label{fig: EffiSkel}
    \vspace{-4mm}
\end{figure*}

As illustrated in Stage 1 of Figure~\ref{fig: EffiSkel}, \textsc{EffiSkel} begins by constructing efficiency-oriented training data from the \textit{APPS} training set. For each natural language description $N$ in \textit{APPS}, we execute multiple corresponding code solutions to measure their runtime performance. These code solutions are then ranked based on execution time, and the top five most efficient ones are retained. During this process, we also remove noisy problems—specifically, problems that lack unit tests or for which no solution passes the provided tests. As a result, we obtain a curated subset of 3,742 problems and 18,710 efficient solutions from the original 5,000 training problems and 117,232 training solutions. This refined dataset, termed \textit{APPS+EFFI}, serves as the foundation for training LLMs to generate efficient code and for extracting efficiency-oriented code skeletons.

\begin{algorithm}[t]
\caption{Skeleton Extraction Methods}
\label{alg:skeleton_extraction}

\KwIn{natural language description $N$; efficient code set $\mathcal{C} = \{C_1, \dots, C_K\}$ solving $N$; reference code $C_{\text{ref}} \in \mathcal{C}$}
\KwOut{skeleton $S_{\text{SMS}}, S_{\text{SSS}} , S_{\text{TAS}}$ of $C_{\text{ref}}$} 

\textcolor{blue}{// SMS: String-based Matching Skeleton}\\
\ForEach{$C_i \in \mathcal{C}$}{
  $T_i \gets \texttt{normalize}(C_i)$ 
}
$L^* \gets \texttt{TCS}(T_1, \dots, T_K)$ , $T_{\text{ref}} \gets \texttt{normalize}(C_{\text{ref}})$ \;
\For{$j = 1$ \KwTo $|T_{\text{ref}}|$}{
  $M_{\text{SMS}}[j] \gets \begin{cases}
    1 & \text{if } T_{\text{ref}}[j] \in L^* \\
    0 & \text{otherwise}
  \end{cases}$
}
$S_{\text{SMS}} \gets \texttt{mask}(C_{\text{ref}}, M_{\text{SMS}})$

\textcolor{blue}{// SSS: Similar Subtree Skeleton}\\
\ForEach{$C_i \in \mathcal{C}$}{
  $\widetilde{T}_i \gets \texttt{normalize\_AST}(\texttt{AST}(C_i))$
}
$S^* \gets \texttt{similar\_subtree}(\widetilde{T}_1, \dots, \widetilde{T}_K)$ , $L_{S^*} \gets \texttt{code\_segments}(S^*, T_{\text{ref}})$ \; 
$S_{\text{SSS}} \gets \texttt{mask}(C_{\text{ref}}, L_{S^*})$

\textcolor{blue}{// TAS: Time-Aware Skeleton}\\
\ForEach{$C_i \in \mathcal{C}$}{
  $\tau_i \gets \texttt{profile}(C_i)$ , $\widetilde{T}_i \gets \texttt{normalize\_AST}(\texttt{AST}(C_i))$
}
$S^* \gets \texttt{similar\_subtree}(\widetilde{T}_1, \dots, \widetilde{T}_K)$ \;
\ForEach{$s \in \bigcup_i \widetilde{T}_i \setminus S^*$}{
  \If{$\bar{\tau}(s) > \delta$}{
     $S^{\text{extra}}_{\text{hot}} \gets S^{\text{extra}}_{\text{hot}} \cup \{s\}$ \;
  }
}
$S^+ \gets S^* \cup S^{\text{extra}}_{\text{hot}}$ , $L_{S^+} \gets \texttt{code\_segments}(S^+, T_{\text{ref}})$ \;
$S_{\text{TAS}} \gets \texttt{mask}(C_{\text{ref}}, L_{S^+})$

\end{algorithm}

\subsection{Skeleton Extraction from Efficient Code}
\label{sec:extract}

To further capture the structural factors that underlie code efficiency, \textsc{EffiSkel} incorporates three complementary skeleton extraction strategies after selecting high-efficiency code  (Stage 2 in Figure~\ref{fig: EffiSkel}). These methods—targeting lexical, syntactic, and dynamic dimensions—are detailed in Algorithm~\ref{alg:skeleton_extraction}. We fill the non-skeleton part with \verb|<MASK>|, which is represented by \verb|_| here.

To extract structurally meaningful and performance-aware skeletons from efficient code, we propose a progressively enhanced skeleton extraction framework. Given a set of functionally correct and efficient code $\mathcal{C} = {C_1, \dots, C_K}$, the goal is to derive a representative structural skeleton from a reference program $C_{\text{ref}}$ by identifying and masking its core regions.

The String-based Matching Skeleton (SMS) focuses on identifying surface-level lexical patterns shared across multiple efficient code. Each code $C_i$ is normalized into a token sequence $T_i$, and the token common subsequence (TCS) $L^*$ is computed across all sequences ${T_1, \dots, T_K}$. The TCS is then projected back to the reference token sequence $T_{\text{ref}}$, generating a binary mask $M_{\text{SMS}}$ and yielding the skeleton $S_{\text{SMS}}$. As shown in the blue-highlighted region of Stage 2 in Figure~\ref{fig: EffiSkel}, the SMS skeleton retains common tokens such as \verb|for| and \verb|return|, reflecting consistent lexical constructs.

The Similar Subtree Skeleton (SSS) captures structural similarity at the syntactic level by leveraging abstract syntax trees (ASTs). Each code $C_i$ is parsed into a normalized AST $\widetilde{T}i$, and the similar subtree $S^*$ is extracted across all $\widetilde{T}i$. Here, \verb|similar_subtree()| uses a relaxed matching strategy based on node types and hierarchical levels; subtrees with a similarity score > 0.8 are treated as “shared”. This shared subtree is then mapped back $T{_\text{ref}}$ to retrieve the corresponding code segments, forming the skeleton $S{_\text{SSS}}$. As shown in the green-highlighted region of Stage 2 in Figure~\ref{fig: EffiSkel}, SSS captures structural alignment even when variable names differ, consistently identifying constructs such as \verb|for-in-zip| loops.

The Time-Aware Skeleton (TAS) enhances SSS by incorporating execution-time profiling information. For each efficient code $C_i$, TAS collects runtime traces by \verb|line_profiler| library inspired by EffiLearner~\cite{huang2024effilearner}. Additionally, any structure outside $S^*$ with average execution time exceeding a threshold defined as the row-average execution time is included as an extra hotspot $S^{\text{extra}}_{\text{hot}}$. The final skeleton $S_{\text{TAS}}$ is constructed by projecting $S^* \cup S^{\text{extra}}_{\text{hot}}$ onto the reference implementation, highlighting both shared and performance-critical regions.
As illustrated in the purple-highlighted region of Stage 2 in Figure~\ref{fig: EffiSkel}, TAS preserves performance-critical statements such as \verb|max()|, which are costly but essential.

Through the above strategies, \textsc{EffiSkel} constructs structure skeleton supervision training samples represented as tuples $(N, C, S, M)$, where $C$ is the efficient code, $S$ is the extracted skeleton, and $M$ is a binary position mask aligned with the skeleton. In ablations, we train one skeleton type per run; our main results use the time-aware skeleton (TAS). These supervisory signals guide the model to focus on performance-critical structures, enabling it to generate code that is both functionally correct and execution-efficient.

\subsection{Model Optimization with Skeleton-Guided Supervision}

To effectively guide the model toward generating functionally correct and execution-efficient code, \textsc{EffiSkel} adopts a multi-task joint training framework, as illustrated in Stage 3 of Figure~\ref{fig: EffiSkel}. This framework incorporates explicit structural supervision extracted from efficient  code, enabling the model to encode generalizable efficiency patterns during training.

Given a natural language description $N$, the model is trained with three forms of supervision: the target efficient code $C$, its corresponding skeleton $S$, and a binary position mask $M$ indicating which tokens in $C$ belong to the skeleton. The training process jointly optimizes both semantic accuracy and structural awareness by integrating these signals into a unified objective.

The primary task focuses on efficient code generation. The model is trained in an autoregressive manner to generate the complete efficient implementation $C$ conditioned on $N$, by minimizing the standard code generation loss $\mathcal L_{\text{code}}$:
\begin{equation}
\mathcal{L}_{\text{code}} = - \sum_{i=1}^{n} \log \mathbb{P}(C_i \mid N, C_{<i})\,.
\end{equation}
The first auxiliary task targets skeleton generation. The model learns to generate the skeleton sequence $S$ from the same input $N$, minimizing the following objective $\mathcal L_{\text{skeleton}}$:
\begin{equation}
\mathcal{L}_{\text{skeleton}} = - \sum_{i=1}^{n} \log \mathbb{P}(S_i \mid N, S_{<i})\,.
\end{equation}
The second auxiliary task provides skeleton-aware token-level supervision. A binary mask $M$ is used to focus learning on performance-critical regions. Only tokens aligned with the skeleton are considered in the loss $\mathcal L_{\text{mask}}$:
\begin{equation}
\mathcal{L}_{\text{mask}} = - \sum_{i=1}^{n} M_i \log \mathbb{P}(C_i \mid N, C_{<i})\,.
\end{equation}
The overall training objective is a weighted combination of all three components, where $\alpha$,$\beta$,$\gamma$ are distinct loss-weight coefficients:
\begin{equation}
\mathcal{L}_{\text{total}} = \alpha \mathcal{L}_{\text{code}} + \beta \mathcal{L}_{\text{skeleton}} + \gamma \mathcal{L}_{\text{mask}}\,.
\end{equation}
This joint optimization framework enables simultaneous improvement in semantic correctness, structural abstraction, and fine-grained structural localization. By integrating explicit structural signals into the learning process, the model is guided to encode generalizable efficiency patterns. The resulting framework balances generation accuracy and structure awareness, providing a unified solution for efficient code generation guided by structured skeleton supervision.

\section{Experimental Design}

\subsection{Research Questions}

We formulate four core research questions (RQs) to evaluate \textsc{EffiSkel}. These questions cover overall performance, robustness under harder benchmarks and stronger baselines, the role of efficient code skeletons and comparisons of fine-tuning strategies.

\begin{longfbox}[margin-top=0pt,margin-bottom=0pt,border-width=0pt,border-left-width=4pt,border-left-color=grey,]
\textit{\textbf{RQ1:} How does \textsc{EffiSkel} perform, compared to existing methods, on specialized benchmarks for efficient code?
}
\end{longfbox}
 

\begin{longfbox}[margin-top=0pt,margin-bottom=0pt,border-width=0pt,border-left-width=4pt,border-left-color=grey,]
\textit{\textbf{RQ2:} Can \textsc{EffiSkel} sustain efficiency gains when evaluated on more challenging benchmarks and against stronger baselines?}
\end{longfbox}


\begin{longfbox}[margin-top=0pt,margin-bottom=0pt,border-width=0pt,border-left-width=4pt,border-left-color=grey,]
\textit{\textbf{RQ3:} What is the impact of different skeleton extraction strategies on model performance?}
\end{longfbox}


\begin{longfbox}[margin-top=0pt,margin-bottom=0pt,border-width=0pt,border-left-width=4pt,border-left-color=grey,]
\textit{\textbf{RQ4:} How does each auxiliary task in \textsc{EffiSkel} affect the final performance?}
\end{longfbox}
 

\subsection{Datasets and Evaluation Metrics}
\label{ER}
We conduct experiments on seven datasets for evaluating the results of efficient code generation, as shown in Table~\ref{tab:table10}:






\begin{table*}[!h]
\centering
\footnotesize
\caption{Summary of seven datasets used for efficient code generation.}
\label{tab:table10}
\begin{threeparttable}
\begin{tabular}{l c c c c c}
\toprule
\textbf{Dataset} & \textbf{Source} & \textbf{Evaluate Tasks} & \textbf{Avg.\ Test Cases}\tnote{$\dagger$} & \textbf{Language} & \textbf{Type} \\
\midrule
Mercury~\cite{du2024mercury}        & LeetCode     & 256     & $+\infty$ & Python & Algorithmic \\
ENAMEL~\cite{qiu2024efficient}     & HumanEval    & 142     & 20        & Python & General-purpose \\
APPS~\cite{hendrycks2021measuring} & CodeForces   & 5{,}000 & 21.2      & Python & Algorithmic \\
EffiBench~\cite{huang2024effibench}& LeetCode     & 1{,}000 & 100       & Python & Algorithmic \\
HumanEval-X (Java)~\cite{zheng2023codegeex} & HumanEval-X  & 163     & 15        & Java   & General-purpose \\
DevEval~\cite{li2024deveval}        & GitHub Repos & 1825      & 2.1       & Python & Repository-level \\
CoderEval~\cite{yu2024codereval} & GitHub Repos & 230      & 10.4       & Python & Repository-level \\
\bottomrule
\end{tabular}
\begin{tablenotes}[para,flushleft]
\footnotesize
\item[$\dagger$] Values denote the average number of test cases per task; Mercury pulls live cases from LeetCode and is effectively unbounded; HumanEval-X extends the original HumanEval benchmark to multiple programming languages, including Java, C++, and Go. CoderEval includes both Python and Java tasks; in this work, we evaluate only the Python subset.
\end{tablenotes}
\end{threeparttable}
\end{table*}

Following previous research, we use the Pass@$k$ metric to measure the functional correctness of code generated by LLMs. To evaluate efficiency, we propose the Efficiency Ratio (ER), a robustness-oriented metric that quantifies the percentage of cases in which one model generates more efficient code than another for the same problem. Although similar in spirit to the ECC metric~\cite{yang2024acecode}, our ER metric generalizes beyond comparisons with the original LLM: $ ER = \frac{1}{N} \sum_{i=1}^{N} {1}(T_i^1 > T_i^2) $
, where N denotes the total number of tasks solved across models. For each task $i$, $T_i^1$ and $T_i^2$ represent the average execution time of the fastest solution generated by different models, respectively. Each execution time $T$ is computed as: $T = \frac{1}{n} \sum_{j=1}^{n} t_j$, where $n$ denotes the number of different input cases, and $t_j$ represents the average execution time over 30 runs under input $j$. The ER is a value between 0 and 1, and is robust to extreme outliers.

However, a limitation of ER is that it only captures the proportion of better-performing solutions, without accounting for the magnitude of the efficiency gains. To address this, we introduce the Average Speedup (AS): $AS = \frac{1}{N} \sum_{i=1}^{N} \frac{T_i^1}{T_i^2}$, where $N$, $T_i^1$, and $T_i^2$ have the same meanings as in the ER metric. An AS value greater than 1 indicates that the model associated with $T^2$ generates more efficient code than the model associated with $T^1$. AS is also pairwise and computed on the same common-correct set by multiple methods to ensure fairness. A larger AS corresponds to greater efficiency improvement. To reduce the impact of heavy-tailed outliers, we winsorize speedup at the 5th–95th percentiles (P5–P95) prior to averaging.

\subsection{Base LLMs and Baselines}

In this paper, we use several LLMs as base models, including Qwen2.5-Coder-1.5B-Instruct~\cite{hui2024qwen2.5coder}, StarCoder2-3B~\cite{lozhkov2024starcoder}, DeepSeek-Coder-6.7B-Instruct~\cite{guo2024deepseek}, CodeLlama-7B-Python-hf~\cite{roziere2023code} and Qwen2.5-Coder-7B-Instruct~\cite{hui2024qwen2.5coder}, all of which have been widely adopted in the domain of efficient code generation~\cite{huangefficoder, ye2025llm4effi, du2024mercury, qiu2024efficient}.

As baselines, we include \textsc{Instruct}\cite{ye2025llm4effi}, \textsc{LLM4EFFI}\cite{ye2025llm4effi} (used here as a prompt-only pipeline that first conducts algorithmic exploration and then implementation optimization, without additional fine-tuning), and two training-based methods. The instruction settings in \textsc{Instruct} are consistent with those in \textsc{LLM4EFFI}. For comparability with our setting, we retrain \textsc{EffiCoder}\cite{huangefficoder} via efficiency-aware supervised fine-tuning on our dataset and train \textsc{CodeDPO}\cite{zhang2024codedpo} via Direct Preference Optimization on preference pairs derived from our dataset; we refer to these tailored versions as \textsc{EffiCoder*} and \textsc{CodeDPO*}.

\subsection{Training and Inference Details}\label{training set}

Due to computational constraints, we adopt parameter-efficient fine-tuning with LoRA, which is widely used in prior work \cite{du2024mercury, paul2025obscuracoder}. For both \textsc{EffiCoder*} and \textsc{EffiSkel}, the model is trained on 18,710 efficient code for 5 epochs with a learning rate of 2e-5, a batch size of 64, and a maximum input sequence length of 1,600. During inference, we use a temperature setting of 0.8 and a top\_p setting of 0.95. For \textsc{CodeDPO*}, we adopt a two-stage preference training: first on 18,710 pairs of correct and incorrect code, then on 4,720 pairs of efficient and inefficient code, with a beta of 0.2 and 800 training steps; inference settings match those above. All experiments are conducted on a cluster equipped with 4 RTX 5880-48GB GPUs.

\section{Experimental Results and Analysis}

\begin{table*}[t]
\centering
\footnotesize
\setlength{\tabcolsep}{5pt}
\caption{Comparison on \textit{Mercury} and \textit{ENAMEL}: Pass@1 (correctness) and ER/AS (efficiency) across five base models under \textsc{Instruct}, \textsc{LLM4EffI}, \textsc{EffiCoder*}, \textsc{CodeDPO*}, and our \textsc{EffiSkel}.}
\label{tab:table1}
\begin{threeparttable}
\begin{tabular}{
l l
S[table-format=2.2] S[table-format=2.2] S[table-format=1.2]
S[table-format=2.2] S[table-format=2.2] S[table-format=1.2]
}
\toprule
& & \multicolumn{3}{c}{\textbf{Mercury}} & \multicolumn{3}{c}{\textbf{ENAMEL}} \\
\cmidrule(lr){3-5}\cmidrule(lr){6-8}
\textbf{Model} & \textbf{Method} & {Pass@1} & {ER (\%)} & {AS} & {Pass@1} & {ER (\%)} & {AS} \\
\midrule
\multirow{5}{*}{Qwen2.5\texttt{-}Coder (1.5B)}
  & \textsc{Instruct}   & 26.17 & 50.00 & 1.00 & 21.13 & 50.00 & 1.00 \\
  & \textsc{LLM4EFFI}   & 28.91 & 50.00 & 1.00 & 23.24 & 50.00 & 1.02 \\
  & \textsc{EffiCoder*} & 39.45 & 54.17 & 1.04 & 30.98 & 55.56 & 1.13 \\
  & \textsc{CodeDPO*}   & 41.02 & 54.17 & 1.13 & 38.73 & 55.56 & 1.22 \\
  & \eskm{\textsc{EffiSkel}} & \eskc{41.41} & \eskc{62.50} & \eskc{1.14} & \eskc{38.73} & \eskc{61.11} & \eskc{1.28} \\
\midrule
\multirow{5}{*}{StarCoder2 (3B)}
  & \textsc{Instruct}   & 11.72 & 50.00 & 1.00 & 7.75  & 50.00 & 1.00 \\
  & \textsc{LLM4EFFI}   & 13.67 & 52.94 & 1.01 & 9.15  & 44.44 & 0.88 \\
  & \textsc{EffiCoder*} & 43.75 & 52.94 & 1.04 & 13.38 & 55.56 & 1.04 \\
  & \textsc{CodeDPO*}   & 43.75 & 58.82 & 1.12 & 17.61 & 55.56 & 1.05 \\
  & \eskm{\textsc{EffiSkel}} & \eskc{47.27} & \eskc{58.82} & \eskc{1.20}
                              & \eskc{20.42} & \eskc{66.67} & \eskc{1.09} \\
\midrule
\multirow{5}{*}{DeepSeek\texttt{-}Coder (6.7B)}
  & \textsc{Instruct}   & 24.61 & 50.00 & 1.00 & 22.54 & 50.00 & 1.00 \\
  & \textsc{LLM4EFFI}   & 27.34 & 51.85 & 1.04 & 23.94 & 52.00 & 1.02 \\
  & \textsc{EffiCoder*} & 70.70 & 55.56 & 1.11 & 39.44 & 52.00 & 1.03 \\
  & \textsc{CodeDPO*}   & 69.92 & 62.96 & 1.25 & 43.66 & 56.00 & 1.11 \\
  & \eskm{\textsc{EffiSkel}} & \eskc{73.05} & \eskc{66.67} & \eskc{1.47} & \eskc{45.77} & \eskc{64.00} & \eskc{1.18} \\
\midrule
\multirow{5}{*}{CodeLlama (7B)}
  & \textsc{Instruct}   & 19.53 & 50.00 & 1.00 & 18.31 & 50.00 & 1.00 \\
  & \textsc{LLM4EFFI}   & 21.48 & 47.83 & 1.02 & 19.01 & 50.00 & 1.00 \\
  & \textsc{EffiCoder*} & 48.05 & 52.17 & 1.07 & 24.65 & 56.25 & 1.06 \\
  & \textsc{CodeDPO*}   & 47.27 & 56.52 & 1.14 & 26.06 & 50.00 & 1.03 \\
  & \eskm{\textsc{EffiSkel}} & \eskc{48.44} & \eskc{65.22} & \eskc{1.30} & \eskc{29.58} & \eskc{68.75} & \eskc{1.24} \\
\midrule
\multirow{5}{*}{Qwen2.5\texttt{-}Coder (7B)}
  & \textsc{Instruct}   & 28.35 & 50.00 & 1.00 & 24.65 & 50.00 & 1.00 \\
  & \textsc{LLM4EFFI}   & 31.25 & 51.61 & 0.94 & 25.35 & 54.55 & 1.01 \\
  & \textsc{EffiCoder*} & 58.59 & 51.61 & 1.02 & 41.55 & 54.55 & 1.01 \\
  & \textsc{CodeDPO*}   & 58.98 & 54.83 & 1.09 & 47.18 & 59.09 & 1.12 \\
  & \eskm{\textsc{EffiSkel}} & \eskc{64.06} & \eskc{58.06} & \eskc{1.16} & \eskc{50.71} & \eskc{63.64} & \eskc{1.27} \\
\bottomrule
\end{tabular}
\end{threeparttable}
\end{table*}

\subsection{RQ1: Overall Performance on \textit{Mercury} and \textit{Enamel}}

We evaluate \textsc{EffiSkel} on two benchmarks using five widely adopted large language models. As shown in Table~\ref{tab:table1}, \textsc{EffiSkel} achieves the \emph{highest efficiency} (ER/AS) across all model–dataset combinations while preserving or improving correctness. For example, on \textit{Mercury} with DeepSeek-Coder (6.7B), it reaches AS \textbf{1.47}, ER \textbf{66.67\%}, and Pass@1 \textbf{73.05}; on \textit{Enamel} with Qwen2.5-Coder (7B), it achieves AS \textbf{1.27}, ER \textbf{63.64\%}, and Pass@1 \textbf{50.71}. StarCoder2 (3B) and CodeLlama (7B) on \textit{Mercury} show similar advantages, indicating sustained efficiency gains without sacrificing Pass@1.

Compared with \textsc{Instruct} and \textsc{LLM4EFFI}, fine-tuning methods encode efficiency patterns into model parameters, enabling single-pass generation that improves ER/AS while retaining correctness characteristics. Among fine-tuning approaches, \textsc{EffiSkel} outperforms single-task efficient code supervised fine-tuning (\textsc{EffiCoder*}) by employing a multi-task objective and an efficiency-guided skeleton (TAS), maintaining consistent gains in functional correctness (Pass@$1$) and execution efficiency (ER, AS). This indicates that incorporating efficient code skeletons into training helps the model learn both functionally correct program structures and runtime-optimized execution patterns, leading to simultaneous improvements in correctness and efficiency. Moreover, preference-based training (\textsc{CodeDPO*}) relies on a global preference between two complete programs, it does not specify which regions should be changed to gain efficiency; in contrast, \textsc{EffiSkel} uses structured skeleton supervision to constrain control flow and data-structure choices, resulting in more robust and generalizable efficiency improvements. Due to space constraints, we provide additional case studies and results in our GitHub repository (See Section ~\ref{github}).

\begin{tcolorbox}[size=title,breakable]
\textit{\textbf{Answer to RQ1:} By combining multi-task learning with efficiency-guided skeletons, \textsc{EffiSkel} improves both functional correctness and execution efficiency on specialized performance benchmarks. Across different models and settings, it attains the highest ER/AS while maintaining competitive or superior Pass@1, demonstrating stable, model-independent improvements beyond prompting-only, single-task SFT, or preference-based methods.}
\end{tcolorbox}

\begin{table*}[t]
\centering
\footnotesize
\setlength{\tabcolsep}{5pt}
\caption{Comparison on \textit{APPS} and \textit{EffiBench}: correctness (Pass@1) and efficiency (ER, AS) across five base models under \textsc{EffiCoder*}, \textsc{CodeDPO*}, and our \textsc{EffiSkel}.}
\label{tab:table2}
\begin{threeparttable}
\begin{tabular}{
l l
S[table-format=1.2] S[table-format=2.2] S[table-format=1.2]
S[table-format=2.2] S[table-format=2.2] S[table-format=1.2]
}
\toprule
& & \multicolumn{3}{c}{\textbf{APPS}} & \multicolumn{3}{c}{\textbf{EffiBench}} \\
\cmidrule(lr){3-5}\cmidrule(lr){6-8}
\textbf{Model} & \textbf{Method} & {Pass@1} & {ER (\%)} & {AS} & {Pass@1} & {ER (\%)} & {AS} \\
\midrule
\multirow{3}{*}{Qwen2.5\texttt{-}Coder (1.5B)}
 & \textsc{EffiCoder*} & 2.94 & 50.00 & 1.00 & 16.80 & 50.00 & 1.00 \\
 & \textsc{CodeDPO*}   & 3.04 & 56.82 & 1.22 & 19.40 & 55.42 & 1.17 \\
 & \eskm{\textsc{EffiSkel}} & \eskc{3.38} & \eskc{62.85} & \eskc{1.31} & \eskc{21.70} & \eskc{60.24} & \eskc{1.24} \\
\midrule
\multirow{3}{*}{StarCoder2 (3B)}
 & \textsc{EffiCoder*} & 3.82 & 50.00 & 1.00 & 13.10 & 50.00 & 1.00 \\
 & \textsc{CodeDPO*}   & 3.84 & 53.44 & 1.04 & 13.70 & 54.55 & 1.01 \\
 & \eskm{\textsc{EffiSkel}} & \eskc{3.90} & \eskc{68.81} & \eskc{1.15} & \eskc{14.70} & \eskc{60.00} & \eskc{1.22} \\
\midrule
\multirow{3}{*}{DeepSeek\texttt{-}Coder (6.7B)}
 & \textsc{EffiCoder*} & 8.02 & 50.00 & 1.00 & 36.10 & 50.00 & 1.00 \\
 & \textsc{CodeDPO*}   & 8.04 & 58.38 & 1.07 & 39.20 & 54.23 & 1.08 \\
 & \eskm{\textsc{EffiSkel}} & \eskc{8.26} & \eskc{71.03} & \eskc{1.17} & \eskc{40.70} & \eskc{61.02} & \eskc{1.20} \\
\midrule
\multirow{3}{*}{CodeLlama (7B)}
 & \textsc{EffiCoder*} & 3.46 & 50.00 & 1.00 & 13.90 & 50.00 & 1.00 \\
 & \textsc{CodeDPO*}   & 3.94 & 58.00 & 1.25 & 15.30 & 56.10 & 1.19 \\
 & \eskm{\textsc{EffiSkel}} & \eskc{4.30} & \eskc{69.61} & \eskc{1.46} & \eskc{16.00} & \eskc{64.63} & \eskc{1.27} \\
\midrule
\multirow{3}{*}{Qwen2.5\texttt{-}Coder (7B)}
 & \textsc{EffiCoder*} & 7.08 & 50.00 & 1.00 & 26.60 & 50.00 & 1.00 \\
 & \textsc{CodeDPO*}   & 7.42 & 60.44 & 1.10 & 28.00 & 56.43 & 1.18 \\
 & \eskm{\textsc{EffiSkel}} & \eskc{7.60} & \eskc{62.79} & \eskc{1.24} & \eskc{28.40} & \eskc{63.37} & \eskc{1.32} \\
\bottomrule
\end{tabular}
\end{threeparttable}
\end{table*}

\subsection{RQ2: Robustness on \textit{APPS} and Transferability on \textit{EffiBench}}

We assess robustness on the harder in-domain \textit{APPS} test split and transferability on cross-benchmark \textit{EffiBench}, comparing against stronger baselines (\textsc{EffiCoder*}, \textsc{CodeDPO*}) under identical training and decoding settings. As summarized in Table~\ref{tab:table2}, \textsc{EffiSkel} consistently delivers higher efficiency (ER/AS) while maintaining or improving correctness (Pass@$1$) across models and datasets on \textit{APPS}. These results indicate that skeleton guidance remains effective under stronger baselines and harder evaluations, and that benefits persist across model scales.

\textsc{EffiSkel} achieves substantial improvements in execution efficiency across all evaluated LLMs on \textit{EffiBench}. Both ER and AS increase consistently, demonstrating that skeleton-guided supervision remains effective even under distributional shift. Additionally, Pass@$1$ also improves across all models indicating that the structural guidance provided by \textsc{EffiSkel} not only enhances runtime performance but also facilitates the generation of semantically more accurate outputs in previously unseen domains.

\begin{tcolorbox}[size=title,breakable]
\textit{\textbf{Answer to RQ2:} \textsc{EffiSkel} maintains its advantage under stronger baselines and harder benchmarks. Across \textit{APPS} and \textit{EffiBench}, it achieves higher ER/AS and matches or improves Pass@$1$, indicating strong in-domain robustness and reliable transfer under distributional shift.}
\end{tcolorbox}

\subsection{RQ3: Effectiveness of Different Efficient Code Skeletons}

To assess the impact of alternative skeleton-extraction strategies, we evaluate three strategy variants (i.e., SMS, SSS, and TAS) across four LLMs. As shown in Table~\ref{tab:table3}, a clear monotonic trend emerges: the richer the structural and semantic cues carried by the skeleton, the higher the efficiency metrics (\textsc{ER}, \textsc{AS}).  TAS consistently delivers the strongest gains, confirming that coupling static structure with runtime profiling provides the most informative supervisory signal.

Although improvements in Pass@$1$ are modest because the skeletons mainly target execution performance rather than correctness, TAS still outperforms the other methods in most cases.  The step-wise rise from SMS to SSS to TAS demonstrates that progressively deeper structure—moving from frequent tokens, to shared AST subtrees, to hot-spot-aware ASTs—translates into proportionally larger speed-ups.  SMS therefore offers an almost zero-cost hint, SSS adds syntactic alignment without requiring execution, and TAS goes further by pinpointing real performance hotpots.

Each skeleton occupies a distinct practical niche.  SMS is purely lexical and remains viable even when neither a parser nor a runtime environment is available; for example, it can provide lightweight efficiency hints for Bash or VBScript snippets embedded in continuous-integration pipelines.  SSS relies only on static ASTs, making it suitable when code execution is prohibited—such as during security-sensitive reviews of C\# micro-services—yet parsers are available; its subtree alignment captures algorithmic skeletons such as loop-nest templates or queue-based BFS patterns. TAS requires instrumentation and profiling, but it pays off in performance-critical workloads; when optimizing Python or Rust numeric kernels, or large-scale ETL jobs, TAS highlights the innermost loops or expensive library calls that static analysis alone may overlook.  This spectrum allows practitioners to select the lightest method that still matches their tooling constraints and optimization goals.

\begin{tcolorbox}[size=title,breakable]
\textit{\textbf{Answer to RQ3:} \textcolor{black}{The richer the skeleton, the greater the efficiency gain.  \underline{SMS} is parser- and runtime-free and thus best for lightweight or parser-scarce settings; \underline{SSS} adds robust static-structural cues wherever ASTs are available; \underline{TAS} couples ASTs with profiling to reveal true hotspots and achieves the largest speed-ups when execution traces can be collected.  Choosing a skeleton therefore hinges on the available analysis budget and the target language environment.}}
\end{tcolorbox}

\begin{table}[t]
\centering
\scriptsize
\caption{Comparison of different efficient code skeletons (i.e., SMS, SSS, and TAS) on efficiency and accuracy metrics across models.
We report deltas row within each model block: $\Delta$P@1 and $\Delta$AS.}
\label{tab:table3}
\begin{threeparttable}
\begin{tabular}{
l l
S[table-format=2.2] S[table-format=+1.2] S[table-format=2.2] S[table-format=1.2] S[table-format=+1.2]
S[table-format=2.2] S[table-format=+1.2] S[table-format=2.2] S[table-format=1.2] S[table-format=+1.2]
}
\toprule
& & \multicolumn{5}{c}{\textbf{Mercury}} & \multicolumn{5}{c}{\textbf{ENAMEL}} \\
\cmidrule(lr){3-7}\cmidrule(lr){8-12}
\textbf{Model} & \textbf{Skeleton}
  & {Pass@1} & {$\Delta$P@1} & {ER (\%)} & {AS} & {$\Delta$AS}
  & {Pass@1} & {$\Delta$P@1} & {ER (\%)} & {AS} & {$\Delta$AS} \\
\midrule
\multirow{4}{*}{Qwen2.5\texttt{-}Coder (1.5B)}
  & w/o & 39.45 & +0.00 & 50.00 & 1.00 & +0.00 & 30.98 & +0.00 & 50.00 & 1.00 & +0.00 \\
  & SMS & 41.02 & +1.57 & 58.06 & 1.13 & +0.13 & 35.92 & +4.94 & 57.14 & 1.04 & +0.04 \\
  & SSS & 39.84 & +0.39 & 61.29 & 1.13 & +0.13 & 37.32 & +6.34 & 57.14 & 1.05 & +0.05 \\
  & TAS & \bfseries 41.41 & \bfseries +1.96 & \bfseries 64.52 & \bfseries 1.15 & \bfseries +0.15
          & \bfseries 38.73 & \bfseries +7.75 & \bfseries 61.90 & \bfseries 1.05 & \bfseries +0.05 \\
\midrule
\multirow{4}{*}{StarCoder2 (3B)}
  & w/o & 43.75 & +0.00 & 50.00 & 1.00 & +0.00 & 13.38 & +0.00 & 50.00 & 1.00 & +0.00 \\
  & SMS & 45.70 & +1.95 & 57.89 & 1.11 & +0.11 & 19.01 & +5.63 & 54.55 & 1.07 & +0.07 \\
  & SSS & 46.48 & +2.73 & 63.16 & 1.14 & +0.14 & 19.01 & +5.63 & 63.63 & 1.05 & +0.05 \\
  & TAS & \bfseries 47.27 & \bfseries +3.52 & \bfseries 63.16 & \bfseries 1.16 & \bfseries +0.16
          & \bfseries 20.42 & \bfseries +7.04 & \bfseries 63.63 & \bfseries 1.10 & \bfseries +0.10 \\
\midrule
\multirow{4}{*}{DeepSeek\texttt{-}Coder (6.7B)}
  & w/o & 70.70 & +0.00 & 50.00 & 1.00 & +0.00 & 39.44 & +0.00 & 50.00 & 1.00 & +0.00 \\
  & SMS & 72.27 & +1.57 & 58.82 & 1.12 & +0.12 & 43.66 & +4.22 & 61.53 & 1.10 & +0.10 \\
  & SSS & 72.27 & +1.57 & 61.76 & 1.16 & +0.16 & 44.37 & +4.93 & 61.53 & 1.13 & +0.13 \\
  & TAS & \bfseries 73.05 & \bfseries +2.35 & \bfseries 64.71 & \bfseries 1.20 & \bfseries +0.20
          & \bfseries 45.77 & \bfseries +6.33 & \bfseries 65.38 & \bfseries 1.21 & \bfseries +0.21 \\
\midrule
\multirow{4}{*}{CodeLlama (7B)}
  & w/o & 48.05 & +0.00 & 50.00 & 1.00 & +0.00 & 24.65 & +0.00 & 50.00 & 1.00 & +0.00 \\
  & SMS & 48.44 & +0.39 & 60.71 & 1.14 & +0.14 & 28.17 & +3.52 & 58.82 & 1.16 & +0.16 \\
  & SSS & 48.05 & +0.00 & 60.71 & 1.15 & +0.15 & 28.87 & +4.22 & 58.82 & 1.21 & +0.21 \\
  & TAS & \bfseries 48.44 & \bfseries +0.39 & \bfseries 64.29 & \bfseries 1.17 & \bfseries +0.17
          & \bfseries 29.58 & \bfseries +4.93 & \bfseries 64.71 & \bfseries 1.26 & \bfseries +0.26 \\
\midrule
\multirow{4}{*}{Qwen2.5\texttt{-}Coder (7B)}
  & w/o & 58.59 & +0.00 & 50.00 & 1.00 & +0.00 & 41.55 & +0.00 & 50.00 & 1.00 & +0.00 \\
  & SMS & 62.11 & +3.52 & 56.25 & 1.19 & +0.19 & 49.30 & +7.75 & 58.33 & 1.12 & +0.12 \\
  & SSS & 62.50 & +3.91 & 59.38 & 1.22 & +0.22 & 48.59 & +7.04 & 61.53 & 1.14 & +0.14 \\
  & TAS & \bfseries 64.06 & \bfseries +5.47 & \bfseries 59.38 & \bfseries 1.25 & \bfseries +0.25
          & \bfseries 50.71 & \bfseries +9.16 & \bfseries 62.50 & \bfseries 1.19 & \bfseries +0.19 \\
\bottomrule
\end{tabular}

\begin{tablenotes}[flushleft]
\item $\Delta$P@1 $= \text{Pass@1} - \text{Pass@1}_{\text{base}}$. $\Delta$AS $= \text{AS} - 1.00$. Best rows per block are bolded.
\end{tablenotes}
\end{threeparttable}

\end{table}

\begin{table}[t]
\centering
\scriptsize
\begin{threeparttable}
\caption{Ablation deltas relative to the \texttt{base} objective ($\mathcal{L}_{\text{code}}$) on \textit{Mercury} and \textit{ENAMEL}. 
\textbf{Config} denotes the training objective variant: \texttt{base} ($\mathcal{L}_{\text{code}}$ only), 
\texttt{+Mask} ($\mathcal{L}_{\text{code}}{+}\mathcal{L}_{\text{mask}}$), 
\texttt{+Skel} ($\mathcal{L}_{\text{code}}{+}\mathcal{L}_{\text{skeleton}}$), and 
\texttt{+Both} ($\mathcal{L}_{\text{code}}{+}\mathcal{L}_{\text{skeleton}}{+}\mathcal{L}_{\text{mask}}$). 
}
\label{tab:table4}
\begin{tabular}{l l
S[table-format=2.2] S[table-format=+1.2] S[table-format=2.2] S[table-format=1.2] S[table-format=+1.2]
S[table-format=2.2] S[table-format=+1.2] S[table-format=2.2] S[table-format=1.2] S[table-format=+1.2]}
\toprule
& & \multicolumn{5}{c}{\textbf{Mercury}} & \multicolumn{5}{c}{\textbf{ENAMEL}} \\
\cmidrule(lr){3-7}\cmidrule(lr){8-12}
\textbf{Model} & \textbf{Config} & {Pass@1} & {$\Delta$P@1} & {ER (\%)} & {AS} & {$\Delta$AS} & {Pass@1} & {$\Delta$P@1} & {ER (\%)} & {AS} & {$\Delta$AS} \\
\midrule
\multirow{4}{*}{Qwen2.5\texttt{-}Coder (1.5B)}
& base   & 39.45 & +0.00 & 50.00 & 1.00 & +0.00 & 30.98 & +0.00 & 50.00 & 1.00 & +0.00 \\
& +Mask  & 39.84 & +0.39 & 55.88 & 1.09 & +0.09 & 35.21 & +4.23 & 52.17 & 1.02 & +0.02 \\
& +Skel  & 39.84 & +0.39 & 58.82 & 1.10 & +0.10 & 36.62 & +5.64 & 56.52 & 1.02 & +0.02 \\
& +Both  & \bfseries 41.41 & \bfseries +1.96 & \bfseries 61.76 & \bfseries 1.14 & \bfseries +0.14
          & \bfseries 38.73 & \bfseries +7.75 & \bfseries 60.89 & \bfseries 1.05 & \bfseries +0.05 \\
\midrule
\multirow{4}{*}{StarCoder2 (3B)}
& base   & 43.75 & +0.00 & 50.00 & 1.00 & +0.00 & 13.38 & +0.00 & 50.00 & 1.00 & +0.00 \\
& +Mask  & 44.14 & +0.39 & 61.90 & 1.07 & +0.07 & 17.61 & +4.23 & 61.54 & 1.05 & +0.05 \\
& +Skel  & 44.92 & +1.17 & 57.14 & 1.13 & +0.13 & 18.31 & +4.93 & 61.54 & 1.04 & +0.04 \\
& +Both  & \bfseries 47.27 & \bfseries +3.52 & \bfseries 66.67 & \bfseries 1.17 & \bfseries +0.17
          & \bfseries 20.42 & \bfseries +7.04 & \bfseries 69.23 & \bfseries 1.12 & \bfseries +0.12 \\
\midrule
\multirow{4}{*}{DeepSeek\texttt{-}Coder (6.7B)}
& base   & 70.70 & +0.00 & 50.00 & 1.00 & +0.00 & 39.44 & +0.00 & 50.00 & 1.00 & +0.00 \\
& +Mask  & 71.09 & +0.39 & 57.14 & 1.06 & +0.06 & 42.25 & +2.81 & 55.56 & 1.08 & +0.08 \\
& +Skel  & 71.48 & +0.78 & 60.00 & 1.09 & +0.09 & 45.07 & +5.63 & 59.26 & 1.10 & +0.10 \\
& +Both  & \bfseries 73.05 & \bfseries +2.35 & \bfseries 62.85 & \bfseries 1.20 & \bfseries +0.20
          & \bfseries 45.77 & \bfseries +6.33 & \bfseries 66.67 & \bfseries 1.20 & \bfseries +0.20 \\
\midrule
\multirow{4}{*}{CodeLlama (7B)}
& base   & 48.05 & +0.00 & 50.00 & 1.00 & +0.00 & 24.65 & +0.00 & 50.00 & 1.00 & +0.00 \\
& +Mask  & 48.05 & +0.00 & 58.62 & 1.11 & +0.11 & 26.06 & +1.41 & 57.89 & 1.15 & +0.15 \\
& +Skel  & 48.44 & +0.39 & 58.62 & 1.12 & +0.12 & 27.46 & +2.81 & 57.89 & 1.20 & +0.20 \\
& +Both  & \bfseries 48.44 & \bfseries +0.39 & \bfseries 62.07 & \bfseries 1.17 & \bfseries +0.17
          & \bfseries 29.58 & \bfseries +4.93 & \bfseries 63.16 & \bfseries 1.24 & \bfseries +0.24 \\
\midrule
\multirow{4}{*}{Qwen2.5\texttt{-}Coder (7B)}
& base   & 58.59 & +0.00 & 50.00 & 1.00 & +0.00 & 41.55 & +0.00 & 50.00 & 1.00 & +0.00 \\
& +Mask  & 62.50 & +3.91 & 58.33 & 1.14 & +0.14 & 47.88 & +6.33 & 57.69 & 1.10 & +0.10 \\
& +Skel  & 63.28 & +4.69 & 61.11 & 1.18 & +0.18 & 48.59 & +7.04 & 61.53 & 1.14 & +0.14 \\
& +Both  & \bfseries 64.06 & \bfseries +5.47 & \bfseries 63.89 & \bfseries 1.25 & \bfseries +0.25
          & \bfseries 50.71 & \bfseries +9.16 & \bfseries 65.38 & \bfseries 1.19 & \bfseries +0.19 \\
\bottomrule
\end{tabular}

\begin{tablenotes}[flushleft]
\item $\Delta$P@1 $= \text{Pass@1} - \text{Pass@1}_{\text{base}}$. $\Delta$AS $= \text{AS} - 1.00$. Best rows per block are bolded.
\end{tablenotes}
\end{threeparttable}
\end{table}

\subsection{RQ4: Ablation Study}
To investigate the effectiveness of different training objectives for \textsc{EffiSkel}, we conduct ablation studies on three components of \textsc{EffiSkel}: efficient code supervision ($\mathcal{L}_{\text{code}}$), skeleton supervision ($\mathcal{L}_{\text{skeleton}}$), and skeleton-aware masking ($\mathcal{L}_{\text{mask}}$), evaluated across four representative LLMs. 

As shown in Table~\ref{tab:table4}, \textsc{EffiSkel}, which combines all three objectives, consistently outperforms the other configurations, highlighting the complementary benefits of the two auxiliary tasks. Both skeleton-aware masking and skeleton supervision individually lead to notable efficiency gains over using code supervision alone, as each introduces structural signals that help guide the model toward more efficient generation. When combined, these objectives yield even greater improvements, indicating that modeling both structural form and structural focus enables more targeted learning of efficiency-relevant code patterns. Their integration enables a more comprehensive learning signal, resulting in more effective generation of efficient code.


In addition, none of the training objectives degrade functional correctness; on the contrary, \textsc{EffiSkel} improves Pass@1 in all cases, indicating that efficiency-oriented supervision can also enhance semantic alignment. We hypothesize that these gains arise from: (i) skeleton supervision steering learning toward performance-critical regions that are central to the program’s core logic, (ii) efficient implementations exhibiting clearer, more explicit control flow, and (iii) a multi-task objective that introduces auxiliary semantic constraints, regularizing the model’s code understanding.

\begin{tcolorbox}[size=title,breakable]
\textit{\textbf{Answer to RQ4:} \textcolor{black}{\textsc{EffiSkel}'s performance gains stem from the complementary effects of skeleton-guided supervision and skeleton-aware token masking. Ablation studies confirm that each component contributes independently to execution efficiency, and their combination via multi-task learning yields the strongest results across models.}}
\end{tcolorbox}

\section{Discussion}

\subsection{What is the impact of the task weighting parameters for \textsc{EffiSkel}?}

As shown in Table~\ref{tab:table5}, we evaluate \textsc{EffiSkel} on Qwen2.5-Coder (1.5B) under different task-weight settings. Adding skeleton signals ($\beta,\gamma>0$) consistently increases efficiency (ER, AS) over code only training ($\alpha=1,\beta=\gamma=0$) while keeping Pass@1 comparable. Performance improves as the objective moves from a single task to a balanced multitask formulation: equal weights ($\alpha=\beta=\gamma=1/3$) already provide a good balance between correctness and efficiency, and giving a slightly larger weight to the code loss ($\alpha=0.6,\beta=0.2,\gamma=0.2$) yields the best ER/AS on both \textit{Mercury} and \textit{Enamel} without reducing Pass@1. Overweighting any single auxiliary term (e.g., $\beta$ or $\gamma$) or reverting to code only training (very large $\alpha$) degrades efficiency and correctness.

\begin{table*}[!h]
\centering
\scriptsize
\caption{Results of ablations with different loss-weight settings on \textit{Mercury} and \textit{ENAMEL}. \textsc{EffiSkel} rows report weight triplets $(\alpha,\beta,\gamma)$ applied to $(\mathcal{L}_{\text{code}},\mathcal{L}_{\text{skeleton}},\mathcal{L}_{\text{mask}})$.}
\label{tab:table5}
\setlength{\tabcolsep}{3pt}
\begin{threeparttable}
\begin{tabular}{
l
S[table-format=1.2] S[table-format=1.2] S[table-format=1.2]
S[table-format=2.2] S[table-format=+1.2] S[table-format=2.2] S[table-format=1.2] S[table-format=+1.2]
S[table-format=2.2] S[table-format=+1.2] S[table-format=2.2] S[table-format=1.2] S[table-format=+1.2]
}
\toprule
& & & & \multicolumn{5}{c}{\textbf{Mercury}} & \multicolumn{5}{c}{\textbf{Enamel}} \\
\cmidrule(lr){5-9}\cmidrule(lr){10-14}
\textbf{Model} & {$\alpha$} & {$\beta$} & {$\gamma$}
& \multicolumn{1}{c}{Pass@1} & \multicolumn{1}{c}{\boldmath$\Delta$P@1} & \multicolumn{1}{c}{ER (\%)} & \multicolumn{1}{c}{AS} & \multicolumn{1}{c}{\boldmath$\Delta$AS}
& \multicolumn{1}{c}{Pass@1} & \multicolumn{1}{c}{\boldmath$\Delta$P@1} & \multicolumn{1}{c}{ER (\%)} & \multicolumn{1}{c}{AS} & \multicolumn{1}{c}{\boldmath$\Delta$AS} \\
\midrule
\multirow{9}{*}{Qwen2.5\texttt{-}Coder (1.5B)}
  & 1.00 & 0.00 & 0.00 & 39.45 & \texttt{+}0.00 & 50.00 & 1.00 & \texttt{+}0.00 & 30.98 & \texttt{+}0.00 & 50.00 & 1.00 & \texttt{+}0.00 \\
  & 0.25 & 0.25 & 0.50 & 38.28 & \texttt{-}1.17 & 49.02 & 0.99 & \texttt{-}0.01 & 30.28 & \texttt{-}0.70 & 50.00 & 1.01 & \texttt{+}0.01 \\
  & 0.50 & 0.00 & 0.50 & 39.84 & \texttt{+}0.39 & 54.90 & 1.09 & \texttt{+}0.09 & 35.21 & \texttt{+}4.23 & 58.33 & 1.02 & \texttt{+}0.02 \\
  & 0.25 & 0.50 & 0.25 & 37.89 & \texttt{-}1.56 & 47.06 & 0.96 & \texttt{-}0.04 & 33.80 & \texttt{+}2.82 & 45.83 & 0.99 & \texttt{-}0.01 \\
  & 0.50 & 0.50 & 0.00 & 39.84 & \texttt{+}0.39 & 58.82 & 1.10 & \texttt{+}0.10 & 36.62 & \texttt{+}4.94 & 54.16 & 1.02 & \texttt{+}0.02 \\
  & 0.50 & 0.25 & 0.25 & 39.84 & \texttt{+}0.39 & 60.78 & 1.12 & \texttt{+}0.12 & 35.91 & \texttt{+}5.64 & 54.16 & 1.03 & \texttt{+}0.03 \\
  & 0.33 & 0.33 & 0.33 & 41.02 & \texttt{+}1.57 & 62.75 & 1.10 & \texttt{+}0.10 & 37.32 & \texttt{+}6.34 & 62.50 & 1.04 & \texttt{+}0.04 \\
  & \best{0.60} & \best{0.20} & \best{0.20} &
    \best{41.41} & \best{+1.96} & \best{66.67} & \best{1.15} & \best{+0.15} &
    \best{38.73} & \best{+7.75} & \best{66.67} & \best{1.05} & \best{+0.05} \\
  & 0.80 & 0.10 & 0.10 & 39.84 & \texttt{+}0.39 & 56.86 & 1.06 & \texttt{+}0.06 & 35.91 & \texttt{+}5.64 & 58.33 & 1.02 & \texttt{+}0.02 \\
\bottomrule
\end{tabular}
\begin{tablenotes}[flushleft]
\item \(\Delta\)P@1 \(=\) Pass@1 \(-\) Pass@1\(_{\text{base}}\); \(\Delta\)AS \(=\) AS \(- 1.00\).
The best-performing weights \((\alpha,\beta,\gamma)=(0.60,\,0.20,\,0.20)\) are bolded and used in experiments.
\end{tablenotes}
\end{threeparttable}
\end{table*}

\subsection{How does model scale influence the effectiveness of \textsc{EffiSkel} in improving execution efficiency?}

\begin{figure}[!t]
  \centering
  \begin{subfigure}{0.48\textwidth}
    \includegraphics[width=\linewidth]{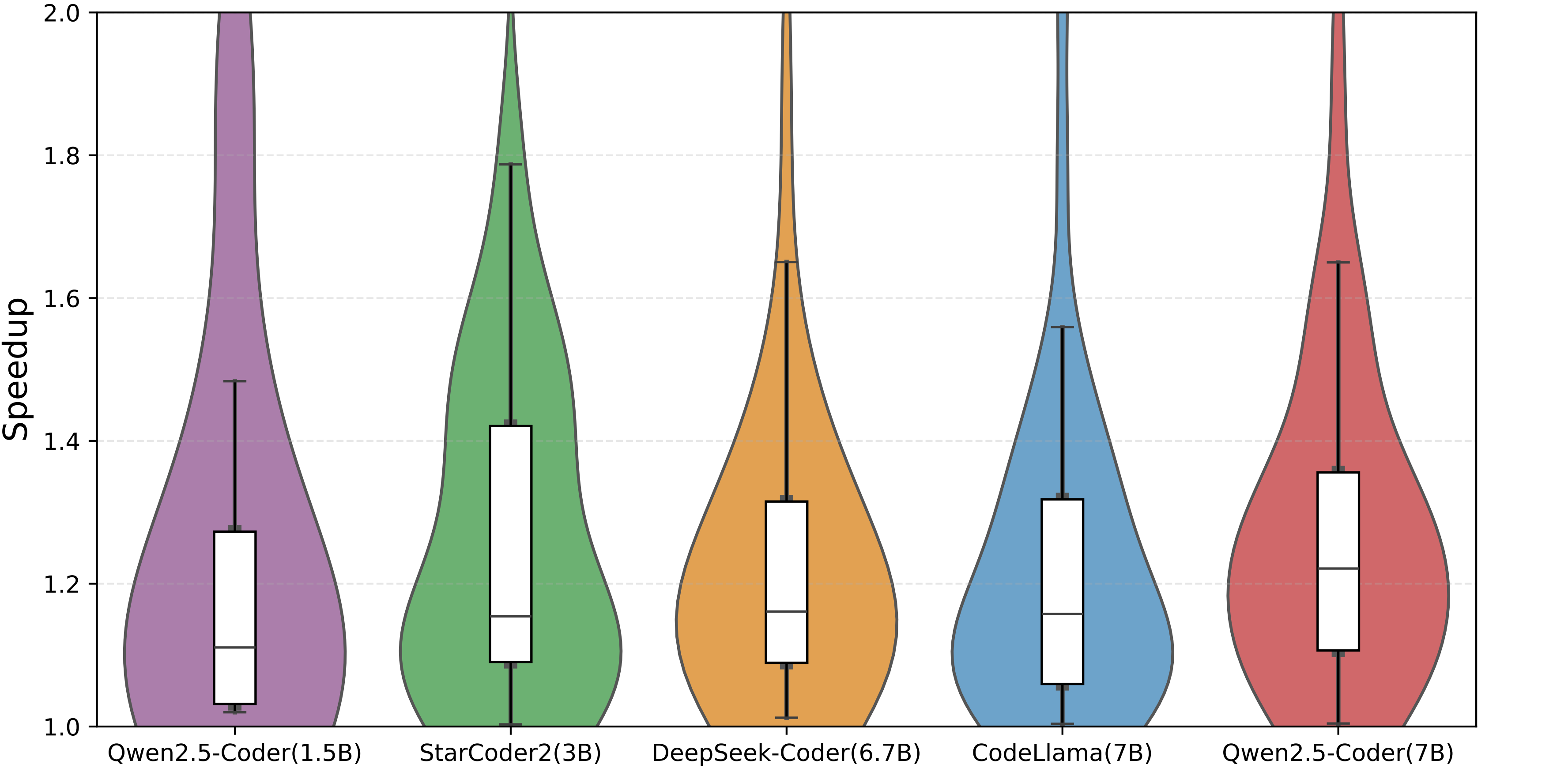}
    \caption{Compare with \textsc{EffiCoder*} on \textit{Mercury}}
  \end{subfigure}
  \hfill
  \begin{subfigure}{0.48\textwidth}
    \includegraphics[width=\linewidth]{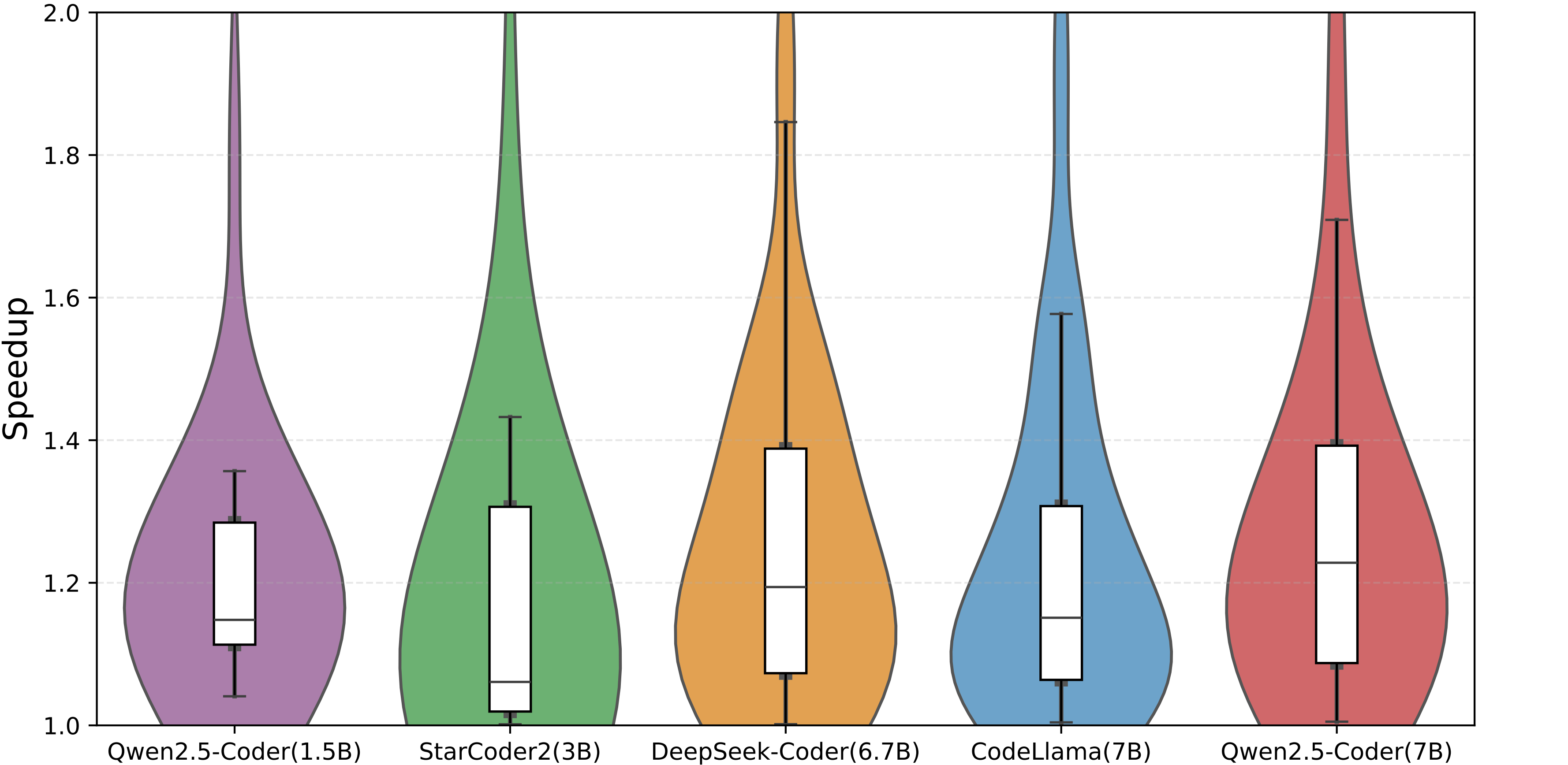}
    \caption{Compare with \textsc{CodeDPO*} on \textit{Mercury}}
  \end{subfigure}

  \par\medskip

  \begin{subfigure}{0.48\textwidth}
    \includegraphics[width=\linewidth]{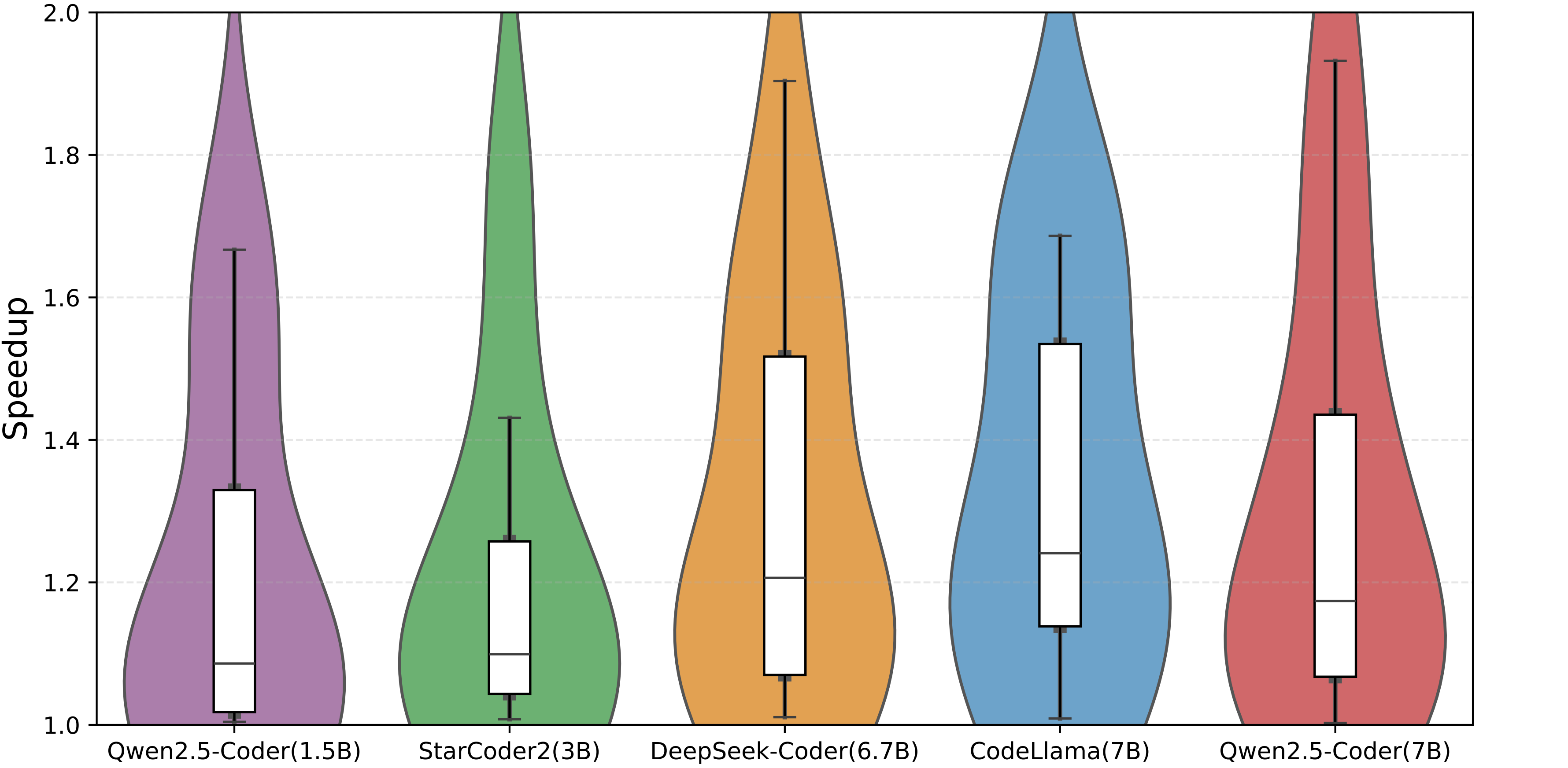}
    \caption{Compare with \textsc{EffiCoder*} on \textit{ENAMEL}}
  \end{subfigure}
  \hfill
  \begin{subfigure}{0.48\textwidth}
    \includegraphics[width=\linewidth]{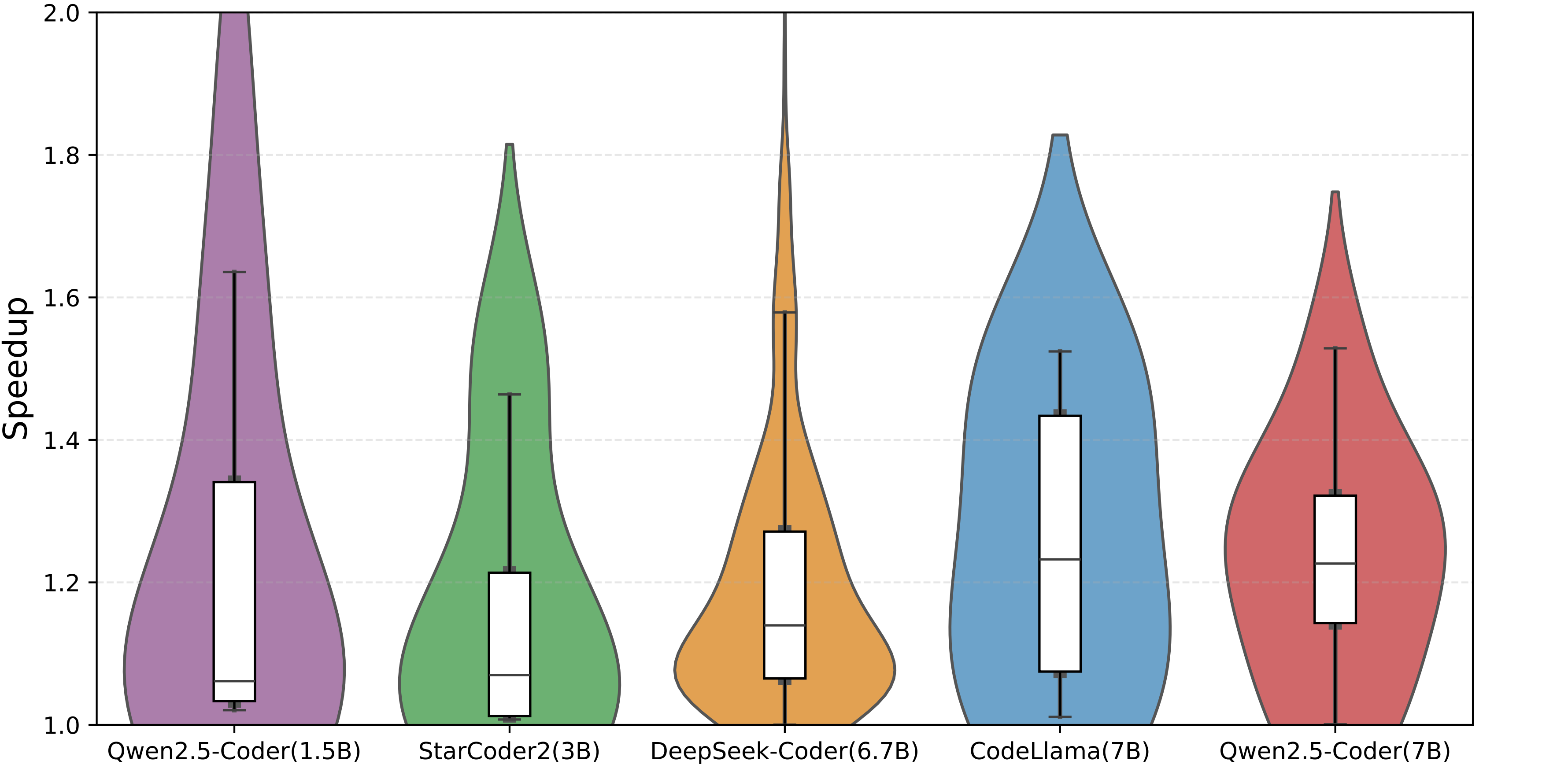}
    \caption{Compare with \textsc{CodeDPO*} on \textit{ENAMEL}}
  \end{subfigure}
  
  \caption{Speedup distributions achieved by \textsc{EffiSkel} on \textit{Mercury} and \textit{ENAMEL}, showing consistent efficiency improvements across benchmarks.}
  \label{fig:speedup-comparison}
\end{figure}

As shown in Fig.~\ref{fig:speedup-comparison}(a–d), on both \textit{Mercury} and \textit{ENAMEL} the speedup in 1.00 to 2.00 distributions shift upward with model size. Larger models — DeepSeek-Coder (6.7B), CodeLlama (7B), and Qwen2.5-Coder (7B) — exhibit higher medians and thicker upper tails than smaller ones — Qwen2.5-Coder (1.5B) and StarCoder2 (3B) — under both baselines (\textsc{EffiCoder*} and \textsc{CodeDPO*}). This pattern indicates that increased capacity enables models to better exploit the efficiency-oriented structural signals encoded by the skeleton, yielding stronger execution-time improvements with comparable dispersion. Smaller models show more modest shifts. Overall, \textsc{EffiSkel} scales favorably: its efficiency gains persist across benchmarks and are amplified as model capacity grows.

\begin{table*}[t]
\centering
\footnotesize
\setlength{\tabcolsep}{4pt}
\caption{Afterburner-inspired iterative optimization at test time (orthogonal to direct efficient code generation). Results on \textit{ENAMEL} with Qwen2.5\mbox{-}Coder (1.5B, 7B) report Pass@1, ER, and AS over three iterations for \textsc{EffiCoder*}, \textsc{CodeDPO*}, and our \textsc{EffiSkel}.}
\label{tab:table6}
\begin{threeparttable}
\begin{tabular}{
l l
S[table-format=2.2] S[table-format=2.2] S[table-format=1.2]
S[table-format=2.2] S[table-format=2.2] S[table-format=1.2]
S[table-format=2.2] S[table-format=2.2] S[table-format=1.2]
}
\toprule
& & \multicolumn{3}{c}{\textbf{Iteration 1}} & \multicolumn{3}{c}{\textbf{Iteration 2}} & \multicolumn{3}{c}{\textbf{Iteration 3}} \\
\cmidrule(lr){3-5}\cmidrule(lr){6-8}\cmidrule(lr){9-11}
\textbf{Model} & \textbf{Method} &
\multicolumn{1}{c}{\textbf{Pass@1}} & \multicolumn{1}{c}{\textbf{ER (\%)}} & \multicolumn{1}{c}{\textbf{AS}} &
\multicolumn{1}{c}{\textbf{Pass@1}} & \multicolumn{1}{c}{\textbf{ER (\%)}} & \multicolumn{1}{c}{\textbf{AS}} &
\multicolumn{1}{c}{\textbf{Pass@1}} & \multicolumn{1}{c}{\textbf{ER (\%)}} & \multicolumn{1}{c}{\textbf{AS}} \\
\midrule
\multirow{3}{*}{Qwen2.5\texttt{-}Coder (1.5B)}
  & \textsc{EffiCoder*} & 33.10 & 50.00 & 1.00 
                        & 33.80 & 50.00 & 1.00 
                        & 35.21 & 50.00 & 1.00 \\
  & \textsc{CodeDPO*}   & 39.44 & 52.94 & 1.01  & 40.85 & 58.82 & 1.04  & 42.96 & 61.76 & 1.06  \\
  & \eskm{\textsc{EffiSkel}} & \eskc{40.14} & \eskc{56.97} & \eskc{1.02}
                        & \eskc{42.25} & \eskc{60.98} & \eskc{1.06}
                        & \eskc{43.66} & \eskc{63.41} & \eskc{1.07} \\
\midrule
\multirow{3}{*}{Qwen2.5\texttt{-}Coder (7B)}
  & \textsc{EffiCoder*} & 42.25 & 50.00 & 1.00  
                        & 43.66 & 50.00 & 1.00   
                        & 45.07 & 50.00 & 1.00   \\
  & \textsc{CodeDPO*}   & 47.89 & 54.76 & 1.04  & 50.00 & 57.16 & 1.05  & 51.41 & 61.90 & 1.05  \\
  & \eskm{\textsc{EffiSkel}} & \eskc{51.40} & \eskc{57.14} & \eskc{1.10}
                        & \eskc{52.11} & \eskc{59.18} & \eskc{1.11}
                        & \eskc{53.52} & \eskc{67.35} & \eskc{1.20} \\
\bottomrule
\end{tabular}
\end{threeparttable}
\end{table*}

\subsection{Does \textsc{EffiSkel} retain its advantage under Afterburner-style iterative optimization, and does a stronger base model amplify the gains?}
\label{iterative optimization}

Across both model sizes and all three iterations in Table~\ref{tab:table6}, \textsc{EffiSkel} consistently attains the highest correctness (Pass@1) and efficiency (ER, AS), outperforming \textsc{EffiCoder} and \textsc{CodeDPO} at every step. Iterative optimization~\cite{du2025afterburner} provides additional improvements, and the Qwen2.5-Coder (7B) variant exhibits larger efficiency gains than the Qwen2.5-Coder (1.5B) model, indicating that skeleton guidance complements iterative optimization and that benefits scale with base-model capacity. Initializing the iterative procedure with \textsc{EffiSkel}-generated programs raises the starting ER/AS and preserves additive gains across rounds, indicating that skeleton guidance strengthens rather than replaces inference-time optimization.

\begin{table*}[!h]
\centering
\footnotesize
\setlength{\tabcolsep}{6pt}
\caption{Comparison on \textit{HumanEval\mbox{-}X (Java)}: correctness (Pass@1) and efficiency (ER/AS)) for \textsc{EffiCoder*}, \textsc{CodeDPO*}, and our \textsc{EffiSkel} on Qwen2.5\mbox{-}Coder (1.5B, 7B).}
\label{tab:table8}
\begin{threeparttable}
\begin{tabular}{
l l
S[table-format=2.2] S[table-format=2.2] S[table-format=1.2]
}
\toprule
\textbf{Model} & \textbf{Method} &
\multicolumn{1}{c}{\textbf{Pass@1}} &
\multicolumn{1}{c}{\textbf{ER (\%)}} &
\multicolumn{1}{c}{\textbf{AS}} \\
\midrule
\multirow{3}{*}{Qwen2.5\texttt{-}Coder (1.5B)}
  & \textsc{EffiCoder*} & 21.34 & 50.00 & 1.00  \\
  & \textsc{CodeDPO*}   & 23.78 & 50.00 & 1.02  \\
  & \eskm{\textsc{EffiSkel}} & \eskc{25.61} & \eskc{71.42} & \eskc{1.07} \\
\midrule
\multirow{3}{*}{Qwen2.5\texttt{-}Coder (7B)}
  & \textsc{EffiCoder*} & 28.05 & 50.00 & 1.00  \\
  & \textsc{CodeDPO*}   & 31.10 & 52.94 & 1.01  \\
  & \eskm{\textsc{EffiSkel}} & \eskc{32.32} & \eskc{60.00} & \eskc{1.07} \\
\bottomrule
\end{tabular}
\end{threeparttable}
\end{table*}

\subsection{How does \textsc{EffiSkel} perform on different programming language?}
\textsc{EffiCoder*} uses efficiency-aware SFT on a curated Java set comprising \textbf{1,177} problems with \textbf{5,885} efficient solutions; \textsc{CodeDPO*} is trained with two-stage DPO—first on \textbf{5,885} correct/incorrect pairs and then on \textbf{1,177} efficient/inefficient pairs. Filtering data on \textit{CodeNet}~\cite{puri2021codenet} and training follows the settings in 
Section~\ref{training set}. As shown in Table~\ref{tab:table8}, \textsc{EffiSkel} outperforms \textsc{EffiCoder*} and \textsc{CodeDPO*} on Java, improving correctness (Pass@1) while also achieving higher efficiency (ER, AS). The gains hold without tuning advantages, indicating that skeleton supervision transfers beyond Python and remains effective under a different syntax and runtime profile.




\subsection{ How does \textsc{EffiSkel} perform on repository-level code generation tasks ?}
\begin{table*}[t]
\centering
\footnotesize
\setlength{\tabcolsep}{5pt}
\caption{Comparison on \textit{DevEval} and \textit{CoderEval}: correctness (Pass@1) and efficiency (ER/AS) across five base models under \textsc{EffiCoder*}, \textsc{CodeDPO*}, and our \textsc{EffiSkel}.}
\label{tab:table11}
\begin{threeparttable}
\begin{tabular}{
l l
S[table-format=2.2] S[table-format=2.2] S[table-format=1.2]
S[table-format=2.2] S[table-format=2.2] S[table-format=1.2]
}
\toprule
& & \multicolumn{3}{c}{\textbf{DevEval}} & \multicolumn{3}{c}{\textbf{CoderEval}} \\
\cmidrule(lr){3-5}\cmidrule(lr){6-8}
\textbf{Model} & \textbf{Method} & {Pass@1} & {ER (\%)} & {AS} & {Pass@1} & {ER (\%)} & {AS} \\
\midrule
\multirow{3}{*}{Qwen2.5\texttt{-}Coder (1.5B)}
 & \textsc{EffiCoder*} &7.23 & 50.00 & 1.00 & 11.30 & 50.00 & 1.00 \\
 & \textsc{CodeDPO*}   &7.67 & 54.76 & 1.05 & 13.47 & 56.25 & 1.01 \\
 & \eskm{\textsc{EffiSkel}} & \eskc{ 7.78} & \eskc{63.10} & \eskc{1.12} & \eskc{15.22} & \eskc{68.75} & \eskc{1.05} \\
\midrule
\multirow{3}{*}{StarCoder2 (3B)}
 & \textsc{EffiCoder*} &6.58 & 50.00 & 1.00 & 10.87 & 50.00 & 1.00 \\
 & \textsc{CodeDPO*}   &7.29 & 55.84 & 1.12 & 12.17 & 58.82 & 1.04 \\
 & \eskm{\textsc{EffiSkel}} & \eskc{ 7.23} & \eskc{62.34} & \eskc{1.20} & \eskc{13.91} & \eskc{70.59} & \eskc{1.06} \\
\midrule
\multirow{3}{*}{DeepSeek\texttt{-}Coder (6.7B)}
 & \textsc{EffiCoder*} & 11.62 & 50.00 & 1.00 & 17.82 & 50.00 & 1.00 \\
 & \textsc{CodeDPO*}   & 12.27 & 52.94 & 1.06 & 19.56 & 58.06 & 1.11 \\
 & \eskm{\textsc{EffiSkel}} & \eskc{12.60} & \eskc{61.44} & \eskc{1.28} & \eskc{21.30} & \eskc{64.52} & \eskc{1.17} \\
\midrule
\multirow{3}{*}{CodeLlama (7B)}
 & \textsc{EffiCoder*} & 10.03 & 50.00 & 1.00 & 13.91 & 50.00 & 1.00 \\
 & \textsc{CodeDPO*}   & 10.41 & 53.97 & 1.14 & 15.65 & 52.63 & 1.09 \\
 & \eskm{\textsc{EffiSkel}} & \eskc{10.74} & \eskc{58.73} & \eskc{1.25} & \eskc{17.39} & \eskc{63.16} & \eskc{1.21} \\
\midrule
\multirow{3}{*}{Qwen2.5\texttt{-}Coder (7B)}
 & \textsc{EffiCoder*} & 11.01 & 50.00 & 1.00 & 18.26 & 50.00 & 1.00 \\
 & \textsc{CodeDPO*}   & 11.45 & 57.93 & 1.17 & 20.87 & 55.88 & 1.12 \\
 & \eskm{\textsc{EffiSkel}} & \eskc{12.49} & \eskc{63.45} & \eskc{1.29} & \eskc{21.74} & \eskc{64.71} & \eskc{1.14} \\
\bottomrule
\end{tabular}
\end{threeparttable}
\end{table*}

To evaluate the effectiveness of \textsc{EffiSkel} in more complex and realistic code generation scenarios, we further conduct experiments on repository-level benchmarks, namely \textit{DevEval} and \textit{CoderEval}. Unlike single-function benchmarks, these datasets require models to generate code within multi-file projects, thereby better reflecting real-world software development settings. We evaluate five base models ranging from 1.5B to 7B parameters and compare \textsc{EffiSkel} against efficiency-oriented baselines, including \textsc{EffiCoder*} and \textsc{CodeDPO*}. Following prior work, we report correctness (Pass@1) and efficiency metrics (ER and AS). Table~\ref{tab:table11} reports the results on \textit{DevEval} and \textit{CoderEval}. Across all five base models, \textsc{EffiSkel} consistently achieves the best performance in terms of both correctness and efficiency on both benchmarks. On \textit{DevEval}, \textsc{EffiSkel} consistently improves efficiency over CodeDPO*, while also achieving higher average speedup across all base models. Similar trends are observed on \textit{CoderEval}, where \textsc{EffiSkel} consistently outperforms both baselines across different model sizes. These results indicate that the efficiency skeletons learned by \textsc{EffiSkel} remain effective in repository-level code generation tasks, and the proposed framework generalizes well to more complex, realistic programming scenarios beyond single-function settings.
\section{Threats to Validity}

We have identified the following two threats to the validity.

\mypara{Effectiveness of Extracted Skeletons} We assume the extracted skeletons—SMS, SSS, and TAS—capture performance-critical structure that guides faster code generation. Because these methods differ in abstraction and fidelity, effectiveness may vary with task complexity and model capacity: SMS is token-level and may generalize poorly, whereas TAS is profiling-driven and more targeted but depends on accurate profiling. Skeleton supervision also hinges on correct alignment between optimized reference code and noise-free runtime measurements; misalignment or profiling noise can weaken the signal. We therefore compare all three extractors; consistent gains suggest that, despite differing granularities, the skeletons provide useful structural guidance.


\mypara{Dataset and Metrics}
\textit{APPS+EFFI} is an efficiency-focused subset of \textit{APPS} with standardized I/O for rigorous, reproducible measurement, though it under-represents real-world heterogeneity.
Additional evaluations are conducted on \textit{Mercury} and \textit{EffiBench}, where consistent gains on human-written \textit{ENAMEL} and \textit{HumanEval\mbox{-}X (Java)} indicate good generalization; we further evaluate \textsc{EffiSkel} on repository-level benchmarks, including \textit{DevEval} and \textit{CoderEval}, to assess its effectiveness in more complex, realistic settings. 
Instead of raw execution time, we report Efficiency Ratio (ER)—the fraction of tasks (among those where across methods are functionally correct) on which \textsc{EffiSkel} runs faster than its comparator—and Average Speedup (AS), computed as the mean per-task speedup of \textsc{EffiSkel} over the comparator on that same set. To reduce noise due to transient hardware load, OS jitter, or outlier inputs, we winsorize the speedup distribution at the 5th and 95th percentiles before averaging; this preserves relative order while dampening extreme values, yielding a more stable estimate of efficiency gains.

\section{Related Work}

\mypara{LLMs for Code} LLMs, such as GPT-4o~\cite{openai2024gpt4o}, DeepSeek-R1~\cite{guo2025deepseek}, Claude 4~\cite{claude4}, and CodeLlama~\cite{roziere2023code}, have demonstrated substantial advancements in a wide range of code-related tasks, including code generation~\cite{wang2025teaching}, code summarization~\cite{ahmed2024automatic}, code completion~\cite{zhang2025hierarchical}, code translation~\cite{yang2024exploring}, and more~\cite{ye2025uncovering, gao2024search}. Among the various code-related tasks, code generation—entailing the automatic transformation of natural language specifications into executable programs—has emerged as a particularly impactful and widely studied application~\cite{sun2024enhancing, gu2023llm, jiang2024survey, huang2024knowledge}. While LLMs have achieved substantial progress in generating functionally correct code, particularly on relatively simple benchmarks such as HumanEval~\cite{chen2021evaluating} and MBPP~\cite{austin2021program}, their ability to produce efficient code has received considerably less attention. This limitation becomes more pronounced in complex or performance-critical tasks, where runtime efficiency are essential. Recent studies~\cite{shi2024efficient, huang2024effibench, niu2024evaluating} have shown that LLM-generated code often lags behind human-written solutions in terms of execution time and memory usage. Moreover, excessive focus on improving the efficiency of code generated by LLMs may come at the expense of their ability to produce functionally correct code~\cite{shehab2024evaluating}. This indicates the need for further research on how to improve the efficiency of LLM-generated code while carefully balancing functional correctness and performance.

\mypara{Efficient Code Generation} The efficiency of generated code by LLMs have attracted significant attention recently~\cite{chambon2025bigobench, qiu2024efficient, du2024mercury, peng2025coffe, coignion2024performance}.Broadly, current approaches fall into two complementary lines: efficient code optimization (improving a draft after it is produced) and direct efficient code generation (training models to emit efficient programs in one shot). 

\textit{Efficient code optimization.} These methods operate at inference time by iteratively refining code with signals from execution or profilers. Effi-Learner~\cite{huang2024effilearner} drives a loop of run–analyze–revise to improve efficiency under correctness constraints. LLM4EFFI~\cite{ye2025llm4effi} decouples logical search from code-level tuning, using targeted edits to improve efficiency while preserving functionality. Afterburner~\cite{du2025afterburner} push this paradigm further with search-and-rewrite cycles guided by runtime feedback and heuristics (e.g., patterned refactorings and micro-optimizations). Such optimization pipelines can yield substantial speedups, but they introduce additional compute and often face trade-offs among readability, maintainability, and guaranteed correctness.

\textit{Direct efficient code generation.} A complementary line aims to make models efficient by construction. Early efforts emphasize efficiency-aware signals—for example, integrating profiling feedback into supervision~\cite{waghjale2024ecco}, decomposing tasks into modules to induce structure~\cite{Pan2025ECode}, or learning from performance-improving edit pairs~\cite{shypula2023learning}; Mercury~\cite{du2024mercury} reports that preference-based training often outperforms plain fine-tuning for efficiency. Code-Optimise~\cite{shypula2023learning} leverages self-generated preference data to align models toward both correctness and efficiency. EffiCoder~\cite{huangefficoder} performs efficiency-aware supervised fine-tuning on curated corpora, while CodeDPO~\cite{zhang2024codedpo} optimizes with preferences over efficient and inefficient code. Collectively, these methods demonstrate consistent gains on efficiency benchmarks.

We follow the direct-generation line but make the structural bias explicit: our method (\textsc{EffiSkel}) introduces efficiency skeletons that capture program-level structure associated with fast execution and uses them as supervision during fine-tuning. This contrasts with optimization pipelines that improve code after generation. We demonstrate this complementarity empirically and discuss integration strategies in Section~\ref{iterative optimization}.

\mypara{Code Generation with Structured Representations} Structural code sketches and syntactic constraints are also used to guide the code generation process~\cite{rabinovich2017abstract, xiong2022l2s, sun2020treegen, sun2019grammar, li2023skcoder}. 
Structure-aware approaches include CodeSAM~\cite{mathai2024codesam}, which injects multi-view guidance via attention masks; CodeS~\cite{zan2024codes}, which decomposes natural-language specifications into sketch-guided subtasks; and GrammarT5~\cite{zhu2024grammart5}, which enforces grammar via labeled constraints and tailored pretraining, improving quality and reducing syntax errors.
Overall, prior structural‐guidance studies center on correctness, readability and syntax, whereas we employ explicit skeleton supervision to elevate runtime efficiency on efficient code generation.

\section{Conclusion}
In this paper, we propose \textsc{EffiSkel}, a sturctured skeleton supervision framework that guides LLMs to generate more efficient code. We design three complementary efficiency skeleton extraction strategies to capture the key structural patterns of efficient code from multiple perspectives and validate their effectiveness across several mainstream code generation models. Experimental results demonstrate that \textsc{EffiSkel} not only improves code execution efficiency but also enhances the model’s ability to encode efficiency-related structural patterns.

\section{Data Availability}
All datasets and source code are publicly available at: \url{https://github.com/YYYY-YuYu/EffiSkel}.
\label{github}

\section*{ACKNOWLEDGMENTS}
This work was supported by the National Natural Science Foundation of China (NSFC) under Grant No. 61602286.


\bibliographystyle{ACM-Reference-Format}
\bibliography{reference}

@STRING{jun = "June"}

@STRING{tosem = "ACM Transactions on Software Engineering and Methodology"}

@article{openai2024gpt4o,
  title={GPT-4o System Card},
  author={Hurst, Aaron and Lerer, Adam and Goucher, Adam P and Perelman, Adam and Ramesh, Aditya and Clark, Aidan and Ostrow, AJ and Welihinda, Akila and Hayes, Alan and Radford, Alec and others},
  journal={arXiv preprint arXiv:2410.21276},
  year={2024}
}

@article{guo2024deepseek,
  title={DeepSeek-Coder: When the Large Language Model Meets Programming--The Rise of Code Intelligence},
  author={Guo, Daya and Zhu, Qihao and Yang, Dejian and Xie, Zhenda and Dong, Kai and Zhang, Wentao and Chen, Guanting and Bi, Xiao and Wu, Yu and Li, YK and others},
  journal={arXiv preprint arXiv:2401.14196},
  year={2024}
}

@article{roziere2023code,
  title={Code llama: Open foundation models for code},
  author={Roziere, Baptiste and Gehring, Jonas and Gloeckle, Fabian and Sootla, Sten and Gat, Itai and Tan, Xiaoqing Ellen and Adi, Yossi and Liu, Jingyu and Sauvestre, Romain and Remez, Tal and others},
  journal={arXiv preprint arXiv:2308.12950},
  year={2023}
}

@inproceedings{niu2024evaluating,
  title={On evaluating the efficiency of source code generated by llms},
  author={Niu, Changan and Zhang, Ting and Li, Chuanyi and Luo, Bin and Ng, Vincent},
  booktitle={Proceedings of the 2024 IEEE/ACM First International Conference on AI Foundation Models and Software Engineering},
  pages={103--107},
  year={2024}
}

@article{cappendijk2024generating,
  title={Generating Energy-efficient code with LLMs},
  author={Cappendijk, Tom and de Reus, Pepijn and Oprescu, Ana},
  journal={arXiv preprint arXiv:2411.10599},
  year={2024}
}

@article{peng2024large,
  title={Large Language Models for Energy-Efficient Code: Emerging Results and Future Directions},
  author={Peng, Huiyun and Gupte, Arjun and Eliopoulos, Nicholas John and Ho, Chien Chou and Mantri, Rishi and Deng, Leo and Jiang, Wenxin and Lu, Yung-Hsiang and L{\"a}ufer, Konstantin and Thiruvathukal, George K and others},
  journal={arXiv preprint arXiv:2410.09241},
  year={2024}
}

@article{shi2024efficient,
  title={Efficient and green large language models for software engineering: Vision and the road ahead},
  author={Shi, Jieke and Yang, Zhou and Lo, David},
  journal={ACM Transactions on Software Engineering and Methodology},
  year={2024},
  publisher={ACM New York, NY}
}

@article{huang2024effilearner,
  title={Effilearner: Enhancing efficiency of generated code via self-optimization},
  author={Huang, Dong and Dai, Jianbo and Weng, Han and Wu, Puzhen and Qing, Yuhao and Cui, Heming and Guo, Zhijiang and Zhang, Jie},
  journal={Advances in Neural Information Processing Systems},
  volume={37},
  pages={84482--84522},
  year={2024}
}

@inproceedings{waghjale2024ecco,
  title={ECCO: Can We Improve Model-Generated Code Efficiency Without Sacrificing Functional Correctness?},
  author={Waghjale, Siddhant and Veerendranath, Vishruth and Wang, Zhiruo and Fried, Daniel},
  booktitle={Proceedings of the 2024 Conference on Empirical Methods in Natural Language Processing},
  pages={15362--15376},
  year={2024}
}

@article{Pan2025ECode,
  title={E-code: Mastering efficient code generation through pretrained models and expert encoder group},
  author={Pan, Yue and Lyu, Chen and Yang, Zhenyu and Li, Lantian and Liu, Qi and Shao, Xiuting},
  journal={Information and Software Technology},
  volume={178},
  pages={107602},
  year={2025},
  publisher={Elsevier}
}

@article{yang2024acecode,
  title={ACECode: A Reinforcement Learning Framework for Aligning Code Efficiency and Correctness in Code Language Models},
  author={Yang, Chengran and Kang, Hong Jin and Shi, Jieke and Lo, David},
  journal={arXiv preprint arXiv:2412.17264},
  year={2024}
}

@article{hendrycks2021measuring,
  title={Measuring coding challenge competence with apps},
  author={Hendrycks, Dan and Basart, Steven and Kadavath, Saurav and Mazeika, Mantas and Arora, Akul and Guo, Ethan and Burns, Collin and Puranik, Samir and He, Horace and Song, Dawn and others},
  journal={arXiv preprint arXiv:2105.09938},
  year={2021}
}

@article{huang2024effibench,
  title={Effibench: Benchmarking the efficiency of automatically generated code},
  author={Huang, Dong and Qing, Yuhao and Shang, Weiyi and Cui, Heming and Zhang, Jie},
  journal={Advances in Neural Information Processing Systems},
  volume={37},
  pages={11506--11544},
  year={2024}
}

@article{hui2024qwen2.5coder,
  title={Qwen2.5-Coder Technical Report},
  author={Hui, Binyuan and Yang, Jian and Cui, Zeyu and Yang, Jiaxi and Liu, Dayiheng and Zhang, Lei and Liu, Tianyu and Zhang, Jiajun and Yu, Bowen and Lu, Keming and Dang, Kai and Fan, Yang and Zhang, Yichang and Yang, An and Men, Rui and others},
  journal={arXiv preprint arXiv:2409.12186},
  year={2024}
}

@article{lozhkov2024starcoder,
  title={Starcoder 2 and the stack v2: The next generation},
  author={Lozhkov, Anton and Li, Raymond and Allal, Loubna Ben and Cassano, Federico and Lamy-Poirier, Joel and Tazi, Nouamane and Tang, Ao and Pykhtar, Dmytro and Liu, Jiawei and Wei, Yuxiang and others},
  journal={arXiv preprint arXiv:2402.19173},
  year={2024}
}

@inproceedings{sun2024enhancing,
  title={Enhancing Code Generation Performance of Smaller Models by Distilling the Reasoning Ability of LLMs},
  author={Sun, Zhihong and Lyu, Chen and Li, Bolun and Wan, Yao and Zhang, Hongyu and Li, Ge and Jin, Zhi},
  booktitle={Proceedings of the 2024 Joint International Conference on Computational Linguistics, Language Resources and Evaluation (LREC-COLING 2024)},
  pages={5878--5895},
  year={2024}
}

@inproceedings{gu2023llm,
  title={Llm-based code generation method for golang compiler testing},
  author={Gu, Qiuhan},
  booktitle={Proceedings of the 31st ACM Joint European Software Engineering Conference and Symposium on the Foundations of Software Engineering},
  pages={2201--2203},
  year={2023}
}

@article{jiang2024survey,
  title={A Survey on Large Language Models for Code Generation},
  author={Jiang, Juyong and Wang, Fan and Shen, Jiasi and Kim, Sungju and Kim, Sunghun},
  journal={arXiv preprint arXiv:2406.00515},
  year={2024}
}

@inproceedings{huang2024knowledge,
  title={Knowledge-aware code generation with large language models},
  author={Huang, Tao and Sun, Zhihong and Jin, Zhi and Li, Ge and Lyu, Chen},
  booktitle={Proceedings of the 32nd IEEE/ACM International Conference on Program Comprehension},
  pages={52--63},
  year={2024}
}

@article{chambon2025bigobench,
  title={BigO(Bench): Can LLMs Generate Code with Controlled Time and Space Complexity?},
  author={Chambon, Pierre and Roziere, Baptiste and Sagot, Benoit and Synnaeve, Gabriel},
  journal={arXiv preprint arXiv:2503.15242},
  year={2025}
}

@article{qiu2024efficient,
  title={How efficient is llm-generated code? a rigorous \& high-standard benchmark},
  author={Qiu, Ruizhong and Zeng, Weiliang Will and Ezick, James and Lott, Christopher and Tong, Hanghang},
  journal={arXiv preprint arXiv:2406.06647},
  year={2024}
}

@article{du2024mercury,
  title={Mercury: A code efficiency benchmark for code large language models},
  author={Du, Mingzhe and Luu, Anh Tuan and Ji, Bin and Liu, Qian and Ng, See-Kiong},
  journal={Advances in Neural Information Processing Systems},
  volume={37},
  pages={16601--16622},
  year={2024}
}

@article{ye2025llm4effi,
  title={LLM4EFFI: Leveraging Large Language Models to Enhance Code Efficiency and Correctness},
  author={Ye, Tong and Huang, Weigang and Zhang, Xuhong and Ma, Tengfei and Liu, Peiyu and Yin, Jianwei and Wang, Wenhai},
  journal={arXiv preprint arXiv:2502.18489},
  year={2025}
}

@inproceedings{wang2021codet5,
  title={CodeT5: Identifier-aware Unified Pre-trained Encoder-Decoder Models for Code Understanding and Generation},
  author={Wang, Yue and Wang, Weishi and Joty, Shafiq and Hoi, Steven C.H.},
  booktitle={Proceedings of the 2021 Conference on Empirical Methods in Natural Language Processing},
  pages={8696--8708},
  year={2021}
}

@article{mathai2024codesam,
  title={CodeSAM: Source Code Representation Learning by Infusing Self-Attention with Multi-Code-View Graphs},
  author={Mathai, Alex and Sedamaki, Kranthi and Das, Debeshee and Mathews, Noble Saji and Tamilselvam, Srikanth and Chimalakonda, Sridhar and Kumar, Atul},
  journal={arXiv preprint arXiv:2411.14611},
  year={2024}
}

@article{zan2024codes,
  title={Codes: Natural language to code repository via multi-layer sketch},
  author={Zan, Daoguang and Yu, Ailun and Liu, Wei and Chen, Dong and Shen, Bo and Li, Wei and Yao, Yafen and Gong, Yongshun and Chen, Xiaolin and Guan, Bei and others},
  journal={arXiv preprint arXiv:2403.16443},
  year={2024}
}

@inproceedings{rabinovich2017abstract,
  title={Abstract Syntax Networks for Code Generation and Semantic Parsing},
  author={Rabinovich, Maxim and Stern, Mitchell and Klein, Dan},
  booktitle={Proceedings of the 55th Annual Meeting of the Association for Computational Linguistics (Volume 1: Long Papers)},
  pages={1139--1149},
  year={2017}
}

@article{xiong2022l2s,
  title={L2S: A framework for synthesizing the most probable program under a specification},
  author={Xiong, Yingfei and Wang, Bo},
  journal={ACM Transactions on Software Engineering and Methodology (TOSEM)},
  volume={31},
  number={3},
  pages={1--45},
  year={2022},
  publisher={ACM New York, NY}
}

@inproceedings{sun2020treegen,
  title={Treegen: A tree-based transformer architecture for code generation},
  author={Sun, Zeyu and Zhu, Qihao and Xiong, Yingfei and Sun, Yican and Mou, Lili and Zhang, Lu},
  booktitle={Proceedings of the AAAI conference on artificial intelligence},
  volume={34},
  number={05},
  pages={8984--8991},
  year={2020}
}

@inproceedings{sun2019grammar,
  title={A grammar-based structural cnn decoder for code generation},
  author={Sun, Zeyu and Zhu, Qihao and Mou, Lili and Xiong, Yingfei and Li, Ge and Zhang, Lu},
  booktitle={Proceedings of the AAAI conference on artificial intelligence},
  volume={33},
  number={01},
  pages={7055--7062},
  year={2019}
}

@inproceedings{zhu2024grammart5,
  title={GrammarT5: Grammar-integrated pretrained encoder-decoder neural model for code},
  author={Zhu, Qihao and Liang, Qingyuan and Sun, Zeyu and Xiong, Yingfei and Zhang, Lu and Cheng, Shengyu},
  booktitle={Proceedings of the IEEE/ACM 46th International Conference on Software Engineering},
  pages={1--13},
  year={2024}
}

@inproceedings{li2023skcoder,
  title={Skcoder: A sketch-based approach for automatic code generation},
  author={Li, Jia and Li, Yongmin and Li, Ge and Jin, Zhi and Hao, Yiyang and Hu, Xing},
  booktitle={2023 IEEE/ACM 45th International Conference on Software Engineering (ICSE)},
  pages={2124--2135},
  year={2023},
  organization={IEEE}
}

@article{guo2025deepseek,
  title={Deepseek-r1: Incentivizing reasoning capability in llms via reinforcement learning},
  author={Guo, Daya and Yang, Dejian and Zhang, Haowei and Song, Junxiao and Zhang, Ruoyu and Xu, Runxin and Zhu, Qihao and Ma, Shirong and Wang, Peiyi and Bi, Xiao and others},
  journal={arXiv preprint arXiv:2501.12948},
  year={2025}
}

@inproceedings{huangefficoder,
  title={EffiCoder: Enhancing Code Generation in Large Language Models through Efficiency-Aware Fine-tuning},
  author={Huang, Dong and Zeng, Guangtao and Dai, Jianbo and Luo, Meng and Weng, Han and QING, Yuhao and Cui, Heming and Guo, Zhijiang and Zhang, Jie},
  booktitle={Forty-second International Conference on Machine Learning},
  year={2025}
}

@article{feng2024llmeffichecker,
  title={Llmeffichecker: Understanding and testing efficiency degradation of large language models},
  author={Feng, Xiaoning and Han, Xiaohong and Chen, Simin and Yang, Wei},
  journal={ACM Transactions on Software Engineering and Methodology},
  volume={33},
  number={7},
  pages={1--38},
  year={2024},
  publisher={ACM New York, NY}
}

@inproceedings{yang2024streamlining,
  title={Streamlining Java Programming: Uncovering Well-Formed Idioms with IdioMine},
  author={Yang, Yanming and Hu, Xing and Xia, Xin and Lo, David and Yang, Xiaohu},
  booktitle={Proceedings of the IEEE/ACM 46th International Conference on Software Engineering},
  pages={1--12},
  year={2024}
}

@inproceedings{song2024revisiting,
  title={Revisiting Code Similarity Evaluation with Abstract Syntax Tree Edit Distance},
  author={Song, Yewei and Lothritz, Cedric and Tang, Xunzhu and Bissyand{\'e}, Tegawend{\'e} and Klein, Jacques},
  booktitle={Proceedings of the 62nd Annual Meeting of the Association for Computational Linguistics (Volume 2: Short Papers)},
  pages={38--46},
  year={2024}
}

@inproceedings{zhao2024easyview,
  title={EasyView: Bringing Performance Profiles into Integrated Development Environments},
  author={Zhao, Qidong and Chabbi, Milind and Liu, Xu},
  booktitle={2024 IEEE/ACM International Symposium on Code Generation and Optimization (CGO)},
  pages={386--398},
  year={2024},
  organization={IEEE}
}

@article{chen2021evaluating,
  title={Evaluating large language models trained on code},
  author={Chen, Mark and Tworek, Jerry and Jun, Heewoo and Yuan, Qiming and Pinto, Henrique Ponde De Oliveira and Kaplan, Jared and Edwards, Harri and Burda, Yuri and Joseph, Nicholas and Brockman, Greg and others},
  journal={arXiv preprint arXiv:2107.03374},
  year={2021}
}

@article{austin2021program,
  title={Program synthesis with large language models},
  author={Austin, Jacob and Odena, Augustus and Nye, Maxwell and Bosma, Maarten and Michalewski, Henryk and Dohan, David and Jiang, Ellen and Cai, Carrie and Terry, Michael and Le, Quoc and others},
  journal={arXiv preprint arXiv:2108.07732},
  year={2021}
}

@article{wang2025teaching,
  title={Teaching code llms to use autocompletion tools in repository-level code generation},
  author={Wang, Chong and Zhang, Jian and Feng, Yebo and Li, Tianlin and Sun, Weisong and Liu, Yang and Peng, Xin},
  journal={ACM Transactions on Software Engineering and Methodology},
  volume={34},
  number={7},
  pages={1--27},
  year={2025},
  publisher={ACM New York, NY}
}

@inproceedings{ahmed2024automatic,
  title={Automatic semantic augmentation of language model prompts (for code summarization)},
  author={Ahmed, Toufique and Pai, Kunal Suresh and Devanbu, Premkumar and Barr, Earl},
  booktitle={Proceedings of the IEEE/ACM 46th international conference on software engineering},
  pages={1--13},
  year={2024}
}

@inproceedings{zhang2025hierarchical,
  title={Hierarchical context pruning: Optimizing real-world code completion with repository-level pretrained code llms},
  author={Zhang, Lei and Li, Yunshui and Li, Jiaming and Xia, Xiaobo and Yang, Jiaxi and Luo, Run and Wang, Minzheng and Chen, Longze and Liu, Junhao and Qu, Qiang and others},
  booktitle={Proceedings of the AAAI Conference on Artificial Intelligence},
  volume={39},
  number={24},
  pages={25886--25894},
  year={2025}
}

@article{yang2024exploring,
  title={Exploring and unleashing the power of large language models in automated code translation},
  author={Yang, Zhen and Liu, Fang and Yu, Zhongxing and Keung, Jacky Wai and Li, Jia and Liu, Shuo and Hong, Yifan and Ma, Xiaoxue and Jin, Zhi and Li, Ge},
  journal={Proceedings of the ACM on Software Engineering},
  volume={1},
  number={FSE},
  pages={1585--1608},
  year={2024},
  publisher={ACM New York, NY, USA}
}

@inproceedings{ye2025uncovering,
  title={Uncovering llm-generated code: A zero-shot synthetic code detector via code rewriting},
  author={Ye, Tong and Du, Yangkai and Ma, Tengfei and Wu, Lingfei and Zhang, Xuhong and Ji, Shouling and Wang, Wenhai},
  booktitle={Proceedings of the AAAI Conference on Artificial Intelligence},
  volume={39},
  number={1},
  pages={968--976},
  year={2025}
}

@inproceedings{gao2024search,
  title={Search-based llms for code optimization},
  author={Gao, Shuzheng and Gao, Cuiyun and Gu, Wenchao and Lyu, Michael},
  booktitle={2025 IEEE/ACM 47th International Conference on Software Engineering (ICSE)},
  pages={254--266},
  year={2024},
  organization={IEEE Computer Society}
}

@inproceedings{shehab2024evaluating,
  title={Evaluating Large Language Models for Code Generation: Assessing Accuracy, Quality, and Performance},
  author={Shehab, Mohammed A and Wardat, Mohammad and Omari, Safwan and Jararweh, Yaser},
  booktitle={2024 2nd International Conference on Foundation and Large Language Models (FLLM)},
  pages={407--416},
  year={2024},
  organization={IEEE}
}

@online{claude4,
  author={Anthropic},
  title={Welcome to Claude 4: Your Partner in AI Innovation},
  url={https://claude4.org/},
  year={2025}
}

@article{zhang2024codedpo,
  title={Codedpo: Aligning code models with self generated and verified source code},
  author={Zhang, Kechi and Li, Ge and Dong, Yihong and Xu, Jingjing and Zhang, Jun and Su, Jing and Liu, Yongfei and Jin, Zhi},
  journal={arXiv preprint arXiv:2410.05605},
  year={2024}
}

@article{paul2025obscuracoder,
  title={ObscuraCoder: Powering Efficient Code LM Pre-Training Via Obfuscation Grounding},
  author={Paul, Indraneil and Yang, Haoyi and Glava{\v{s}}, Goran and Kersting, Kristian and Gurevych, Iryna},
  journal={arXiv preprint arXiv:2504.00019},
  year={2025}
}

@article{du2025afterburner,
  title={Afterburner: Reinforcement Learning Facilitates Self-Improving Code Efficiency Optimization},
  author={Du, Mingzhe and Tuan, Luu Anh and Liu, Yue and Qing, Yuhao and Huang, Dong and He, Xinyi and Liu, Qian and Ma, Zejun and Ng, See-kiong},
  journal={arXiv preprint arXiv:2505.23387},
  year={2025}
}

@article{shypula2023learning,
  title={Learning performance-improving code edits},
  author={Shypula, Alexander and Madaan, Aman and Zeng, Yimeng and Alon, Uri and Gardner, Jacob and Hashemi, Milad and Neubig, Graham and Ranganathan, Parthasarathy and Bastani, Osbert and Yazdanbakhsh, Amir},
  journal={arXiv preprint arXiv:2302.07867},
  year={2023}
}

@article{peng2025coffe,
  title={Coffe: A code efficiency benchmark for code generation},
  author={Peng, Yun and Wan, Jun and Li, Yichen and Ren, Xiaoxue},
  journal={Proceedings of the ACM on Software Engineering},
  volume={2},
  number={FSE},
  pages={242--265},
  year={2025},
  publisher={ACM New York, NY, USA}
}

@inproceedings{coignion2024performance,
  title={A performance study of llm-generated code on leetcode},
  author={Coignion, Tristan and Quinton, Cl{\'e}ment and Rouvoy, Romain},
  booktitle={Proceedings of the 28th international conference on evaluation and assessment in software engineering},
  pages={79--89},
  year={2024}
}

@article{liu2024evaluating,
  title={Evaluating language models for efficient code generation},
  author={Liu, Jiawei and Xie, Songrun and Wang, Junhao and Wei, Yuxiang and Ding, Yifeng and Zhang, Lingming},
  journal={arXiv preprint arXiv:2408.06450},
  year={2024}
}

@inproceedings{zheng2023codegeex,
  title={Codegeex: A pre-trained model for code generation with multilingual benchmarking on humaneval-x},
  author={Zheng, Qinkai and Xia, Xiao and Zou, Xu and Dong, Yuxiao and Wang, Shan and Xue, Yufei and Shen, Lei and Wang, Zihan and Wang, Andi and Li, Yang and others},
  booktitle={Proceedings of the 29th ACM SIGKDD Conference on Knowledge Discovery and Data Mining},
  pages={5673--5684},
  year={2023}
}

@article{puri2021codenet,
  title={Codenet: A large-scale ai for code dataset for learning a diversity of coding tasks},
  author={Puri, Ruchir and Kung, David S and Janssen, Geert and Zhang, Wei and Domeniconi, Giacomo and Zolotov, Vladimir and Dolby, Julian and Chen, Jie and Choudhury, Mihir and Decker, Lindsey and others},
  journal={arXiv preprint arXiv:2105.12655},
  year={2021}
}

@inproceedings{li2024deveval,
  title={DevEval: A manually-annotated code generation benchmark aligned with real-world code repositories},
  author={Li, Jia and Li, Ge and Zhao, Yunfei and Li, Yongmin and Liu, Huanyu and Zhu, Hao and Wang, Lecheng and others},
  booktitle={Findings of the Association for Computational Linguistics: ACL 2024},
  pages={3603--3614},
  year={2024},
  publisher={Association for Computational Linguistics}
}

@inproceedings{yu2024codereval,
  title={CodeEval: A benchmark of pragmatic code generation with generative pre-trained models},
  author={Yu, Hailong and Shen, Bo and Ran, Dong and Zhang, Jian and Zhang, Qi and Ma, Yifan and others},
  booktitle={Proceedings of the 46th IEEE/ACM International Conference on Software Engineering},
  pages={1--12},
  year={2024},
  publisher={IEEE/ACM}
}

@String{Computer = "{IEEE} Computer" }

\end{document}